\documentclass[prb,aps,superscriptaddress,floatfix,tightenlines,showpacs,twocolumn]{revtex4-2}
\usepackage{graphicx}
\usepackage{dcolumn}   
\usepackage{bm}        
\usepackage{amssymb}   
\usepackage{amsmath}
\usepackage{url}
\usepackage{layout}
\usepackage{color}
\usepackage{epsfig}
\usepackage{graphicx}
\usepackage{booktabs}
\usepackage{float}
\usepackage[unicode=true,pdfusetitle,bookmarks=true,bookmarksnumbered=false,bookmarksopen=false,breaklinks=false,pdfborder={0 0 1},backref=false,colorlinks=false]
{hyperref}
\hypersetup{colorlinks,citecolor=blue,linkcolor=blue,filecolor=magenta,urlcolor=blue}

\begin{document}

\title{Unconventional magnetism in spin-orbit coupled systems}
\author{Jian-Keng Yuan}
\affiliation{Fudan University, Shanghai 200433, China}
\affiliation{New Cornerstone Science Laboratory, Department of Physics, School of Science, Westlake University, Hangzhou 310024, Zhejiang, China}
\author{Zhiming Pan}
\email{panzhiming@xmu.edu.cn}
\affiliation{Department of Physics, Xiamen University, Xiamen 361005, China}
\author{Congjun Wu}
\email{wucongjun@westlake.edu.cn}
\affiliation{New Cornerstone Science Laboratory, Department of Physics, School of Science, Westlake University, Hangzhou 310024, Zhejiang, China}
\affiliation{Institute for Theoretical Sciences, Westlake University, Hangzhou 310024, Zhejiang, China}
\affiliation{Key Laboratory for Quantum Materials of Zhejiang Province, School of Science, Westlake University, Hangzhou 310024, Zhejiang, China}
\affiliation{Institute of Natural Sciences, Westlake Institute for Advanced Study, Hangzhou 310024, Zhejiang, China}

\date{\today}

\begin{abstract}
``Unconventional magnetism" was proposed to describe the exotic states arising from Landau-Pomeranchuk instabilities in the spin channel nearly two decades ago.
Its odd-partial-wave-channel (e.g. $p$-wave) states break parity giving rise to the dynamic generation of spin-orbit coupling, 
while its even-partial-wave-channel (e.g. $d$-wave) states break time-reversal symmetry. 
Both types of states can exhibit collinear and non-collinear spin configurations over Fermi surfaces with 
the former and latter termed as the $\alpha$ and $\beta$-phases, respectively. 
The collinear states in even partial-wave channels are in the same symmetry class of ``altermagnetism".
In this work, we investigate unconventional magnetism in both $p$- and $d$-wave channels within spin-orbit coupled systems with parity and time-reversal symmetries maintained. 
Based on the Ginzburg-Landau free energy analysis, the $p$-wave channel yields the gyrotropic, Rashba, Dresselhaus-type spin-orbit couplings.
They compete and mix evolving from the $\beta$-phase to the $\alpha$-phase with various types of spin-momentum lockings.
Analyses are performed in parallel for the $d$-wave unconventional magnetism.
We emphasize that the single-particle dispersion is not sufficient to justify the spin-group type symmetry of the full Hamiltonian. 
Furthermore, Goldstone manifolds and excitations are examined in each unconventional magnetic phase. 
\end{abstract}
\maketitle

\section{Introduction}
Itinerant ferromagnetism breaks rotational symmetry in the spin channel but not in the orbital channel. 
The resulting spin polarization is uniform when moving the electron around the Fermi surface.
This situation is similar to conventional $s$-wave superconductors in which the phase of the gap function does not change over the Fermi surface. 
In other words, it could be deemed as the ``$s$-wave magnetism".
Furthermore, there exist unconventional superconductivity and pairing superfluid in higher partial-wave channels.
Over two decades ago, in analogy to triplet fermion pairing superfluidity, such as $^3$He-A and B-phases, Wu and Zhang studied spin-channel Pomeranchuk instabilities which spontaneously generate spin-orbit orderings in the $p$-wave channel of spin-$\frac{1}{2}$ systems and the $d$-wave channel of spin-$\frac{3}{2}$ cases \cite{wu2004soc}.
Furthermore, the Pomeranchuk instability in spin channels was analyzed in general partial-wave channels by Wu, Sun, Fradkin, and Zhang \cite{wu2007fermi}, yielding spin-multipole ordering over Fermi surfaces, which later was termed as ``unconventional magnetism"~\cite{wu07website}.
Similarly to unconventional superconductivity, unconventional magnetic orders form non-trivial representations of the rotational group. 
Unconventional magnetism was later extended to orbital band systems as a candidate mechanism for explaining the nematic metamagnetic state observed in Sr$_3$Ru$_2$O$_7$ \cite{lee2009}.
Unconventional magnetism under magnetic dipolar interactions, which is a form of spin-orbit coupled interaction, was also investigated \cite{li2012}.

The Landau-Fermi liquid theory provides a natural microscopic mechanism for unconventional magnetism \cite{wu2004soc,wu2007fermi,li2012soc}.
When the Landau interaction parameter in the $F^a_l$ channel with $l\ge 1$ is negatively large, {\it i.e.}, $F^a_l<-(2l+1)$ in 3D and $F^a_l<-2$ in 2D
\cite {pomeranchuk1958stability}, the Fermi surface instability occurs leading to unconventional magnetism. 
Unconventional magnetism can be classified by the evenness of the partial-wave number $l$, and whether the spin texture over the Fermi surface is collinear or non-collinear. 
In analogy to $^3$He-A and B phases, the collinear and non-collinear unconventional magnetic states are termed as $\alpha$- and $\beta$-phases, respectively \cite{wu2004soc}.
For odd values of $l$, the unconventional magnetic states remain time-reversal (TR) invariant but break parity (P).
In contrast, for even values of $l$, such states break TR but maintain P.
In the $\alpha$-phase, the Fermi surfaces exhibit an anisotropic distortion, while those in the $\beta$-phase remain circular or spherical and undistorted.
The $\alpha$-phase of the $p$-wave state was studied by Hirsch under the name of the spin-split state ~\cite{Hirsch90p6820, Hirsch90p6828}. 
The $d$-wave collinear phase was also speculated upon by Oganesyan {\it et. al.} \cite{oganesyan2001quantum} under the name of ``nematic-spin-nematic" state, exhibiting Fermi surface distortions of orthogonal ellipses with opposite spin components.

The recent research focus on the ``altermagnetism" is related to a particular subset of unconventional magnetism -- the $\alpha$-phase in the even partial wave channels \cite{smejkal2022emerging,smejkal2022beyond,liu2022spin,xiao2024spin,chen2024enumeration,jiang2024enumeration}.
They are connected to each other via the adiabatic evolution without changing the symmetry. 
Due to the breaking of TR symmetry, the Bloch-wave from the periodic lattice potential can be viewed as superposed by plane waves with the same spin but opposite wavevectors such that the real space spin textures can develop respecting the $d$-wave spin-group type symmetry. 
Recent experiments in altermagnets have uncovered phenomena such as the anomalous Hall effect \cite{feng2022anomalous,sato2024altermagnetic,attias2024intrinsic} and Kramers degeneracy lifting \cite{lee2024broken,krempasky2024altermagnetic,reimers2024direct}. 
Key to this evolution is the interplay between the electron band structure and lattice magnetic orders. 
The classification of spin-space groups has systematically organized magnetic orders with negligible spin-orbit coupling (SOC) \cite{liu2022spin,xiao2024spin,chen2024enumeration,jiang2024enumeration}, where spin and orbital rotational symmetries are independent.

In this paper, we study the unconventional magnetism in the presence of explicit SOC using the Ginzburg-Landau (GL) formalism.
We use the $p$-wave magnetism in 3D and the $d$-wave case in 2D as representative examples. 
The independent spin $SO(3)_S$ and orbital $SO(3)_L$ rotational symmetries are explicitly broken down to $SO(3)_J$ symmetry in the total angular momentum channel.   
At the quadratic level of the GL analysis, three distinct regimes are identified in $p$-wave channel
corresponding to the scalar, vector, and tensor representations of the $SO(3)_{J}$ group, respectively.
The quartic terms in the GL free energy mix phases in different channels, and phase boundaries are determined. 
A parallel classification is extended to the 2D $d$-wave case, corresponding to representations associated with the total angular momentum $J_z = 1,2,3$, respectively. 
Furthermore, the topological defects and Goldstone modes in each phase are elucidated.
Although the single-particle dispersion shows features reminiscent of spin-group type symmetry, the full Hamiltonian and the many-body physics do not exhibit such a symmetry.

The rest part of this paper is organized as follows.
In Sect.~\ref{sect:GL}, we introduce the symmetry analysis and Ginzburg-Landau framework for unconventional magnetism in odd/even partial-wave channels.
Sect.~\ref{sect:pwave} focuses on the 3D $p$-wave unconventional magnetism in the spin-orbit coupled systems, characterizing three distinct phases through their spin-textures and symmetry-breaking features.
Sect.~\ref{sect:dwave} investigates the $d$-wave unconventional magnetism under spin-orbit coupling. 
In Sect.~\ref{sect:dis}, the relation between SOC and spin-group type symmetry is discussed. 
Sect.~\ref{sect:con} summarizes the results on spin-orbit coupled unconventional magnetism.

\section{Unconventional magnetism}
\label{sect:GL}
In this section, we review the symmetry analysis of the unconventional states in both odd and even partial wave channels in the absence of 
the external spin-orbit coupling
\cite{wu2004soc,wu2007fermi}. 

\subsection{Symmetry and Ginzburg-Landau analysis}
\label{sect:GL1}

In the $l$-th partial-wave channel, the order parameters for unconventional magnetism are defined as follows
\begin{align}
\hat{Q}_{\mu b}(\vec{r}) =\psi^{\dagger}(\vec{r}) \sigma_{\mu} g_b^{(l)}(-i\hat{\nabla}) \psi(\vec{r}),
\end{align}
where Pauli matrices $\sigma_{\mu}$ ($\mu=x,y,z$) describe the orientation in the spin space; 
$g_b^{(l)}$ is the $l$-th partial-wave symmetric tensor in the orbital channel with $b=1,2$ in 2D and $b=1,2,..., 2l+1$ in 3D; 
$\hat{\nabla}$ is defined as an unit operator as $\hat{\nabla}=\vec{\nabla}/|\vec{\nabla}|$. 
In the ordered phase, the order parameters exhibit non-vanishing expectation values as
\begin{eqnarray}
n_{\mu b}\equiv \langle \hat{Q}_{\mu b}(\vec{r})\rangle
=\int \frac{d^d \mathbf{k}}{(2\pi)^d} \langle \psi^\dagger(\mathbf{k}) 
\sigma_\mu \psi(\mathbf{k}) \rangle g_b^{(l)}(\hat{\mathbf{k}}),
\end{eqnarray}
with unit vector $\hat{\mathbf{k}}=\mathbf{k}/|\mathbf{k}|$.
This leads to a dynamic generation of unconventional magnetism due to electron correlations \cite{wu2004soc,wu2007fermi}.
In other words, $n_{\mu b}$ is a spin multipole order defined in momentum space.

The effective Hamiltonian for electrons in the $l$-wave channel in the $d$-dimension is 
\begin{equation}
H = \int \frac{d^{d}\mathbf{k}}{(2\pi)^{d}} \psi^{\dagger}(\mathbf{k}) \big[\epsilon(\mathbf{k})-\mu+\sigma_{\mu} n_{\mu b} g_{b}^{(l)}(\hat{\mathbf{k}})\big]
\psi(\mathbf{k}),
\end{equation}
where $n_{\mu b}$ induces an effective ``magnetic" field $m_{\mu}(\hat{\mathbf{k}}) =n_{\mu b} g_b^{(l)}(\hat{\mathbf{k}})$ in momentum space.

The unconventional magnetic states exhibit different symmetry patterns in the odd and even partial-wave channels.
Under TR transformation $T$ and inversion (parity) transformation $P$, the order parameter transforms in the way,
\begin{align}
T n_{\mu b} T^{-1} =(-1)^{l+1} n_{\mu b},\quad
P n_{\mu b} P^{-1} =(-1)^{l} n_{\mu b},
\end{align}
corresponding to the $Z_2^T$ TR and the $Z_2^P$ inversion symmetries, respectively.
The $p$-wave case, as a representative of odd partial waves, maintains TR symmetry and breaks parity, 
playing the role of an effective spin-orbit coupling. 
In contrast, the $d$-wave case, as a representative of even partial waves, breaks TR symmetry but keeps parity invariant.
Upon turning the lattice potentials, there should not be spin textures for the $p$-wave unconventional magnetic state due to TR symmetry. 
On the contrary, the Bloch waves for the $d$-wave case can develop spin textures within the unit cell, hence, it yields the same symmetry class of the recently proposed ``altermagnetic" states \cite{smejkal2022beyond,smejkal2022emerging}. 
As explained in Ref.~\cite{wu2007fermi}, unconventional magnetic states often exhibit symmetries of a combined spatial and spin rotations not necessarily at the same rotation angles, which are termed ``spin-group" symmetries in later literature \cite{liu2022spin,xiao2024spin,chen2024enumeration,jiang2024enumeration}.

In addition to TR and parity, the order parameter $n_{\mu b}$ reflects several kinds of symmetries. 
For convenience, we use the 3D case as an example, and the situation for 2D is quite similar.
Under the independent rotational transformations in the spin space ($R_S\in SO(3)_S$) and orbit space ($R_L\in SO(3)_L$), 
tensor $n_{\mu b}$ transforms in the following way,
\begin{align}
n_{\mu b} \rightarrow R_{S,\mu\nu} n_{\nu a} R_{L,ab}^{-1}.
\end{align}

Up to the quartic level, the Ginzburg-Landau free energy, respecting the discrete TR and inversion symmetries, as well as the $SO(3)_S\otimes SO(3)_L$ symmetry, 
takes the form,
\begin{align}
F[n] =\alpha \mathrm{Tr}(n n^t) +\beta_1 [\mathrm{Tr}(n n^t)]^2 +\beta_2 \mathrm{Tr}(n n^t)^2,
\label{eq:free}
\end{align}
with $n^t$ represents the transpose of $n$.
Here, $\alpha$, $\beta_1$ and $\beta_2$ are interaction parameters that can be determined by Landau parameters and density of states, whose expressions are given in Ref.~\cite{wu2007fermi}.
$n n^t$ is a $3\times 3$ matrix. 
As shown in Appendix \ref{appendixGL3d}, the eigenvalues of $n n^t$ are positive definite which are parameterized as $f_i^2$ with $i=1,2,3$.
Then 
\begin{eqnarray}
F[f_i^2] =\alpha\sum_i f_i^2
+\beta_1 \big(\sum_i f_i^2\big)^2
+\beta_2 \sum_i f_i^4.
\end{eqnarray}
On the condition that  $\mathrm{Tr}(n n^t)$
is fixed to $\lambda^2$, i.e., $f_1^2+f_2^2+f_3^2=\lambda^2$,
the minimization of Eq.~(\ref{eq:free}) yields
\begin{eqnarray}
f_1^2=f_2^2=f_3^2=\frac{1}{3}\lambda^2,
\end{eqnarray}
at $\beta_2>0$, and without loss of generality we have
\begin{eqnarray}
f_1^2=\lambda^2, \ \ \, f_2^2=f_3^2=0,
\end{eqnarray}
at $\beta_2<0$ \cite{wu2004soc,wu2007fermi}.

\subsection{The $\alpha$ and $\beta$-phases}
In the rest part of this article, we examine
two representative situations.
First, we address the 3D $p$-wave case in the angular momentum channel $l=1$, a configuration intrinsically linked to spin-orbit coupling phenomena. 
Subsequently, we extend our analysis to 
2D systems for the general $l$-th partial-wave channels, with a focus on the $d$-wave ($l=2$) symmetry that establishes connections with the emerging paradigm of altermagnetism.

We first consider the $p$-wave channel in the 3D case.
The 3D $\alpha$-phase or $\beta$-phase is favored at $\beta_2<0$ and $\beta_2>0$,
respectively \cite{wu2004soc,wu2007fermi}.
If $\beta_2<0$, the minimization of Eq.~(\ref{eq:free}) yields the order parameter configuration $n_{\mu b}=\bar{n}\hat{d}_{\mu}\hat{e}_b$ where $\hat{d}$ and $\hat{e}$ are two unit vectors in the spin and orbital space, respectively.
Without loss of generality, we choose $\hat{d}=\hat{e}=\hat z$, such that the order parameter is given by $n_{\mu b}=\bar{n}\delta_{\mu z} \delta_{bz}$.
The dispersions of the two spin orientations split and the original degenerate Fermi surfaces undergo opposite-direction distortions for opposite spin polarizations.
The residue symmetry in the $\alpha$-phase is the orbital channel $SO(2)_L$ rotation around $\hat e$,  the spin channel $SO(2)_S$ rotation around $\hat d$, and the combined spin-orbit $Z_2$ rotation around $\hat e \times \hat d$ at the angle of $\pi$ such that $\hat e\to -\hat e$, and $\hat d \to -\hat d$.
The resulting Goldstone manifold is 
\begin{eqnarray}
&&[SO(3)_L \otimes SO(3)_S]/[SO(2)_L \otimes SO(2)_S \ltimes Z_2]\nonumber \\
&=&(S^2_L \otimes S^2_S )/Z_2.
\end{eqnarray}
The associated Goldstone modes are two channels of orbital-waves and two channels of spin-dipole wave. 

On the contrary, for $\beta_{2}>0$, minimizing the free energy favors the configuration $n_{\mu b}=\bar{n}D_{\mu b}$, where $D_{\mu b}$ is an arbitrary $SO(3)$ matrix. 
While maintaining an isotropic Fermi surface, this phase develops a non-trivial spin texture around the Fermi surface, which is known as the $\beta$-phase \cite{wu2004soc,wu2007fermi}.
The residue symmetry in such a phase is the overall spin-orbit rotation $SO(3)_{J}$. 
If in the absence of the external spin-orbit coupling, the Goldstone manifold corresponding to the $\beta$-phase is 
\begin{eqnarray}
[SO(3)_{L}\otimes SO(3)_S]/SO(3)_{J}=SO(3).
\label{eq:GS_beta}
\end{eqnarray}
The associated Goldstone modes are three branches of spin-orbit waves corresponding to relative spin-orbit rotations. 

\begin{figure}[t]
\centering
\includegraphics[width=0.8\linewidth]{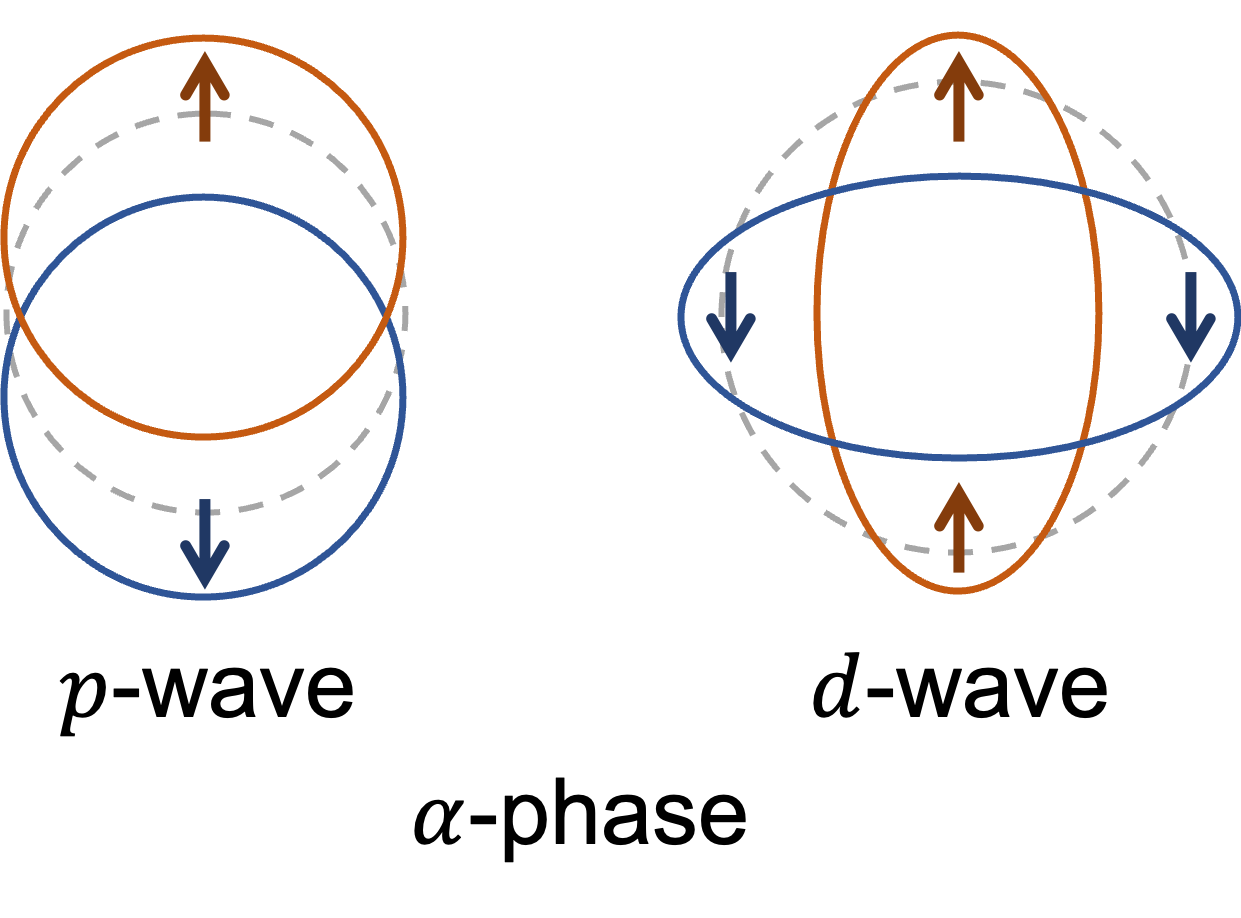}
\caption{Fermi surface distortions in the $\alpha$ phase ($\beta_2<0$) for the $p$-wave channel ($l=1$) and the $d$-wave channel ($l=2$).}
\label{fig:alphaphase}
\end{figure}

In the 2D cases, it will be convenient to introduce two vectors in the spin space, $\vec{n}_1=n_{\mu 1}$ and $\vec{n}_2=n_{\mu 2}$, and formulate the free energy in the form \cite{wu2007fermi},
\begin{equation}
\begin{aligned}
&F[\vec{n}_1,\vec{n}_2]
=\alpha(|\vec{n}_1|^2 +|\vec{n}_2|^2)  \\
&+(\beta_{1}+\beta_2)
\big(|\vec{n}_{1}|^{2} +|\vec{n}_{2}|^{2}\big)^{2} 
-2\beta_{2}
|\vec{n}_{1}\times\vec{n}_{2}|^{2}.
\end{aligned}
\label{eq:GL2d}
\end{equation}
The relevant $l$-th partial-wave tensors are given by, 
\begin{equation}
g_{1}^{(l)}(\mathbf{k})=\cos(l\theta_{\mathbf{k}}),\quad
g_{2}^{(l)}(\mathbf{k})=\sin(l\theta_{\mathbf{k}})
\end{equation}
where $\theta_{\mathbf{k}}$ is the azimuthal angle of $\mathbf{k}$.

If $\beta_2<0$, the minimization of the free energy Eq.~(\ref{eq:GL2d}) favors $\vec{n}_1\parallel \vec{n}_2$.
Without loss of generality, we can consider $\vec{n}_1=\bar{n}\hat{z}$ with vanishing $\vec{n}_2$, leading to the effective $l$-th partial-wave spin-orbital coupling: 
\begin{equation}
H_{d}= \bar{n} \cos(l\theta_{\mathbf{k}}) \sigma_{z}.
\end{equation}
The resulting $\alpha$-state exhibits Fermi surface splitting for the opposite spin orientation, as depicted in Fig.~(\ref{fig:alphaphase}) .
Of particular contemporary interest is the $l=2$ partial-wave ($d$-wave) channel realization in this phase, which manifests as a momentum-dependent spin-splitting of the Fermi surface, a hallmark feature now recognized as “altermagnetism” \cite{smejkal2022beyond,smejkal2022emerging}. The residual symmetry in the $\alpha$-phase is the spin $SO(2)_S$ rotation around $\hat n_1$, the orbital rotation at the angle of $2\pi/l$, and the spin-group type symmetry which rotate at the angle of $\pi/l$ around combined with spin-rotation at the angle $\pi$ around the $\hat n_1$-axis. 
Hence, the Goldstone manifold is 
\begin{eqnarray}
&&[SO(2)_L \times SO(3)_S ]/[SO(2)_S \ltimes Z_2 \times Z_l] \nonumber \\
&=& 
[SO(2)_L/Z_l] \otimes [S^2_S/Z_2].
\end{eqnarray}
The associated Goldstone modes are one channel of orbital-waves and two channels of spin-dipole wave. 

\begin{figure}[t]
\centering
\includegraphics[width=0.9 \linewidth]{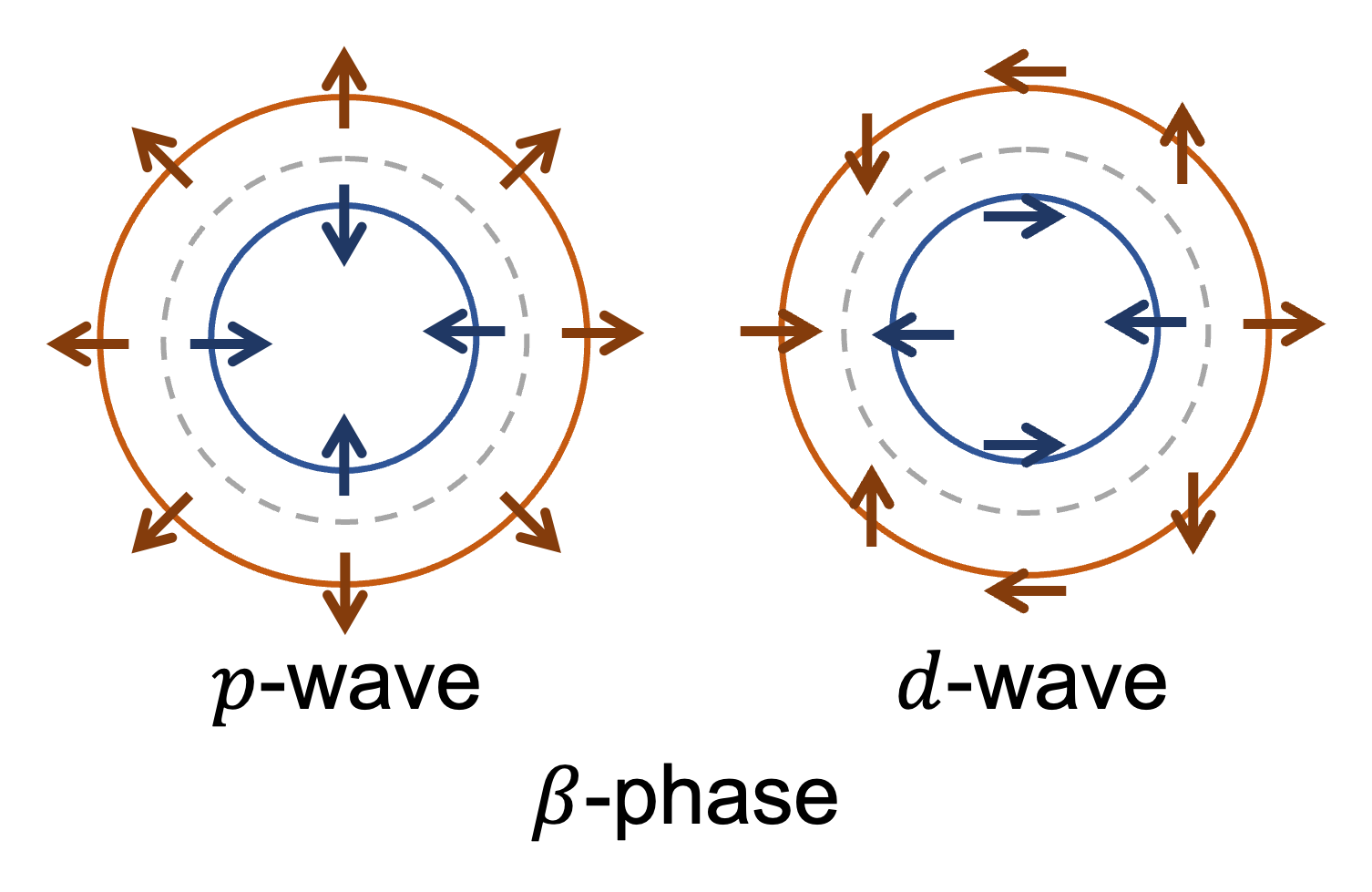}
\caption{Fermi surface and spin-texture in the $\beta$-phase ($\beta_2>0$) for the $p$-wave channel ($l=1$) and the 
$d$-wave channel ($l=2$).
}
\label{fig:betaphase}
\end{figure}

On the other hand, for $\beta_2>0$, the minimization of the free energy in Eq.~(\ref{eq:GL2d}) favors $\vec{n}_1\perp \vec{n}_2$ with equal amplitudes.
For simplicity, we can assume $\vec{n}_{1}=\bar{n} \hat{x}$ and $\vec{n}_{2}=\bar  n\hat{y}$, which leads to the Hamiltonian
\begin{equation}
H_{d}= \bar{n} \big( \cos(l\theta_{\mathbf{k}}) \sigma_{x}+\sin(l\theta_{\mathbf{k}}) \sigma_{y}\big).
\end{equation}
The isotropic dispersion exhibits spin-texture around the Fermi surface, which is characterized by the 2D winding number $l$ for each channel, as depicted in Fig.~(\ref{fig:betaphase}).
The $l=1$ $\beta$-phase typically leads to the generation of the Rashba and Dresselhaus spin-orbit couplings \cite{wu2004soc,wu2007fermi}.
The residual symmetry in this phase is generated by $J^\prime_z=L_z + l \sigma_z /2 $.
Hence, the Goldstone manifold is
\begin{eqnarray}
[SO(2)_L \otimes SO(3)_S]/SO(2)_{J^\prime} =
SO(3).
\end{eqnarray}
The associated Goldstone modes are also three branches of spin-orbit waves.

\section{The $p$-wave unconventional magnetism under 
spin-orbit coupling}
\label{sect:pwave}

In the presence of the external spin-orbit coupling, the full rotational symmetry within the spin and orbital channels, the $SO(3)_S\otimes SO(3)_L$ symmetry of the Fermi liquid, 
breaks down to the overall rotational symmetry of $SO(3)_J$ where $J=S+L$ denotes the total angular momentum \cite{wu2004soc,wu2007fermi,li2012soc}.
In this section, we study the consequence of  unconventional magnetism in the $p$-wave channel in 3D under the external SOC. 

\subsection{Decomposition of order parameters}
The order parameter $n_{\mu b}$ in the $p$-wave channel carries the symmetry of the spin current. 
It can be decomposed into the following sectors according to the SO(3)$_J$ symmetry, i.e.,  $\mathbf{3}\otimes \mathbf{3}=\mathbf{1}\oplus \mathbf{3}\oplus \mathbf{5}$,
as
\begin{eqnarray}
n= T+ A +S,
\end{eqnarray}
with each tensor given by,
\begin{eqnarray}
T&=&\frac{\mathrm{Tr}(n)}{3}; \ \ \ 
A=\frac{n-n^t}{2};  \ \ \
S=\frac{n+n^t}{2}-T. 
\end{eqnarray}
Here, $T$ is the identity matrix multiplied by the trace of $n_{\mu b}$; 
$A$ is the antisymmetric part, and $S$ is the traceless symmetric part of $n_{\mu b}$, respectively.
They transform according to the $J=0,1,2$ irreducible representations of $SO(3)_J$, respectively.
In other words, they are the scalar, vector, and
rank-2 spherical tensor, respectively. 
In the presence of the external spin-orbit coupling, the remaining symmetry is $SO(3)_{J}$, and the quadratic term 
of the GL free energy splits as
\begin{align}
F_2[n] =\alpha_0 \mathrm{Tr}(T T^t) +\alpha_1\mathrm{Tr}(A A^t) +\alpha_2\mathrm{Tr}(S S^t).	
\label{eq:GLF2nSOC}
\end{align}
Each channel of $J=0,1,2$ may develop different ordering patterns explained as follows. 
In the case where $\alpha_0=\alpha_1=\alpha_2=\alpha$, Eq.~(\ref{eq:GLF2nSOC}) reduces to the quadratic term of the GL free energy of Eq.~(\ref{eq:free}).

\begin{figure}[t]
\centering
\includegraphics[width=1.0\linewidth]{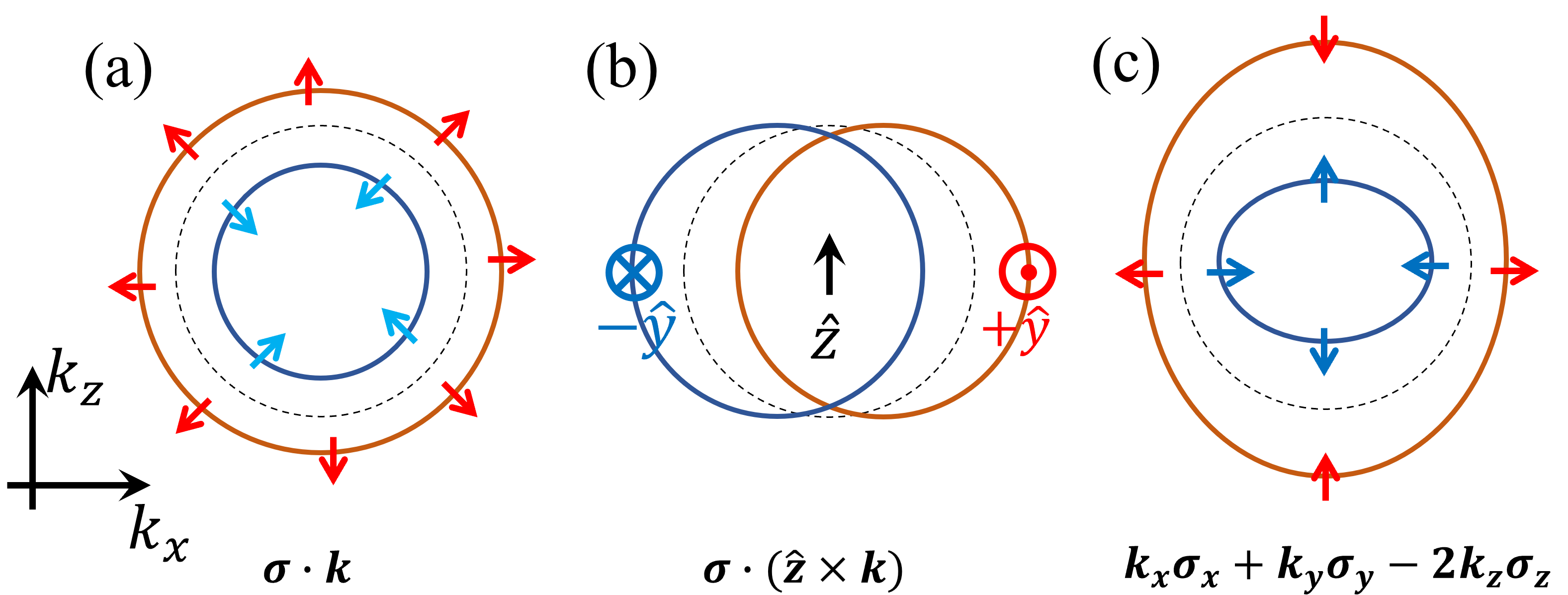}
\caption{Schematic illustration of spin textures on the Fermi surface cross-section in the $k_x$-$k_z$ plane for (a) gyrotropic, (b) ferroelectric (Rashba) and (c) multipolar (Dresselhaus) phase. 
The original Fermi surfaces (dotted circle) are splitting into two Fermi surfaces (solid line) with spins of different directions. }
\label{fig:soctexture}
\end{figure}

The order parameter corresponding to $J=0$ is a pseudo-scalar such that $n_{\mu b}=\bar{n} \delta_{\mu b}$.
It was termed as the gyrotropic phase in the transition-metal dichalcogenide semimetal $1$T-TiSe$_2$ \cite{xu2020gyrotropic,ueda2021correlation}, and it is a special case of the $\beta$-phase.
The effective spin-orbit coupling takes the following form:
\begin{align}
H_{\text{gyro}}
=\bar{n}  \mathbf{k} \cdot \vec{\sigma} ,
\label{eqSOC:Gyrot}
\end{align}
exhibiting the full $SO(3)_{J}$ symmetry.
Taking the eigenvalues $\pm 1$ of the helicity operator $\vec{\sigma} \cdot \mathbf{\hat k}$, the energy dispersion relations are $\xi_{\pm}(\mathbf{k})=\epsilon(\mathbf{k}) -\mu \pm \bar{n} k$.
The corresponding spin texture exhibits a monopole configuration in momentum space, as shown in Fig.~(\ref{fig:soctexture}).
The 3D winding numbers of the spin polarization over the two Fermi surfaces exhibiting opposite helicities are:
\begin{eqnarray}
w=\frac{1}{4\pi}\oint d^2 \mathbf{\hat k} 
~ \hat n(\mathbf{\hat k}) \cdot \left(\partial_\mu \hat n (\mathbf{\hat k}) \times \partial_\nu \hat n (\mathbf{\hat k}) \right)
=\pm 1,
\end{eqnarray}
respectively, where $\hat n(\mathbf{k})=\pm \hat{\mathbf{k}}$.
The spin current pattern consistent with the $J=0$ sector is that the spin polarization is parallel to the direction of the spatial flow. 
Now, with explicit spin-orbit symmetry breaking, such a phase only further breaks parity.
The originally gapless modes
associated with the Goldstone manifold Eq.~(\ref{eq:GS_beta}) become gapped.

The order parameter matrix of the $J=1$ channel is skew-symmetric, $n^t=-n$, which can be mapped onto a pseudo-vector $\bar{n} \hat{u}$.
Such an order can occur as a Pomeranchuk instability in fermion systems with magnetic dipolar interactions \cite{li2012}, which represent a form of spin-orbit coupled interaction explicitly breaking
the $SO(3)_L\otimes SO(3)_S$ symmetry.
The effective spin-orbit coupling takes the Rashba form,
\begin{align}
H_{\text{R}}
= \bar{n}  \vec \sigma \cdot ( \mathbf{\hat{u}} \times \mathbf{k}),	\label{eqSOC:Ferro}
\end{align}
with the unit vector $\hat{u}$.
The residual symmetry of the $J=1$ channel order is $SO(2)_J$, corresponding to the rotation around the $\hat{u}$ axis.
For the case that $\hat{u}=\hat{z}$, 
\begin{align}
H_{\text{R}}
= \bar{n}  \big( k_x\sigma_y -k_y\sigma_x \big).
\end{align}
The associated Goldstone manifold is:
\begin{eqnarray}
SO(3)_J/SO(2)_J=S^2,    
\end{eqnarray}
giving rise to two Goldstone modes. 
Eq.~(\ref{eqSOC:Ferro}) exhibits the same symmetry as the
Rashba SOC, nevertheless, for the case of a 3D Fermi surface the SOC disappears along the direction where
$\mathbf{k}\parallel \hat u$.
In momentum space, the spin-momentum locking exhibits
a vortex loop configuration with a winding number $w=1$.

The order parameter in the $J=2$ sector reduces to the 5D space of a traceless symmetric matrix $S_{\mu b}=S_{b\mu}$.
The effective SOC takes the Dresselhaus form:
\begin{align}
H_{\text{D}}
= \frac{1}{2} S_{\mu b} (\sigma_{\mu} \hat{\mathbf{k}}_b +\sigma_{b} \hat{\mathbf{k}}_{\mu}).	
\label{eqSOC:Multi}
\end{align}
For the case where
$S_{\mu b}=\frac{\bar{n}}{2\sqrt{3}}\mbox{diag}(1,1,-2)$, 
\begin{align}
H_{\text{D}}
= \frac{\bar{n}}{2\sqrt{3}} 
\left( k_x \sigma_x +k_y \sigma_y
-2k_z \sigma_z
\right).
\end{align}
The corresponding energy dispersion relations are $\xi_{\pm}(\mathbf{k})=\epsilon(\mathbf{k}) -\mu \pm \frac{\bar{n}}{2\sqrt{3}} \sqrt{k^2 +3k_z^2}$, and the 3D winding number over the Fermi surface 
$w=\frac{1}{4\pi}\oint d^2 \hat k ~ \hat n(\mathbf{\hat k}) \cdot (\partial_\mu \hat n (\mathbf{\hat k}) \times \partial_\nu \hat n (\mathbf{\hat k}))
=\mp 1$, respectively, 
where $n(\mathbf{k})=\pm (k_x, k_y, -2k_z)$.
For the diagonal order with $J=2, J_z=0$,
the residual symmetry is $D(\infty)_J$, containing the rotation around the $z$-axis and a $\pi$-angle rotation around any axis in the $xy$-plane. 
The Goldstone manifold is:
\begin{eqnarray}
SO(3)_J/D(\infty)_J= S^2/Z_2. 
\end{eqnarray}

There also exist four other typical spin-momentum locking texture configurations exhibiting effective SOCs, such as 
\begin{eqnarray}
&&\hat{\mathbf{k}}_x \sigma_y +\hat{\mathbf{k}}_y \sigma_x, \ \ \, 
\hat{\mathbf{k}}_x \sigma_x -\hat{\mathbf{k}}_y \sigma_y, \nonumber \\
&& \hat{\mathbf{k}}_y \sigma_z +\hat{\mathbf{k}}_z \sigma_y, \ \ \,
\hat{\mathbf{k}}_z \sigma_x + \hat{\mathbf{k}}_x \sigma_z.
\end{eqnarray}
They exhibit anti-vortex loop configurations with a 2D winding number
$w=-1$. 
In these phases, the residual symmetry is discrete, the $D_{4d}$ symmetry with respect to a plane specified by the order, say, the $k_x$-$k_y$-plane for the $\hat k_x \sigma_y +\hat k_y \sigma_x$-type order.

\subsection{The effect of the quartic $\beta_2$ terms}
In principle, in a spin-orbit coupled GL free energy, there exist 3 independent quadratic terms for the gyrotropic ($J=0$), Rashba ($J=1$), and multipolar ($J=2$) orders as shown in Eq.~(\ref{eq:GLF2nSOC}).
For simplicity, we only consider the case that $\alpha_1=\alpha_2=\alpha$, and denote $\alpha_0=\alpha +\Delta \alpha_0$.
Then the GL free energy at the quadratic level becomes:
\begin{equation}
\begin{aligned}
F[n]  = \alpha \mathrm{Tr} (n n^t)+
 \frac{\Delta \alpha_0}{3} (\mathrm{Tr}n)^2.
\end{aligned}
\end{equation}
Such a term breaks the degeneracy of the phase with $J=0$ from the other two phases.
The phase with $J=0$ is disfavored and favored at $\Delta \alpha_0>0$ and $\Delta \alpha_0<0$,
respectively.

The quartic $\beta_1$-term $\beta_1 \mathrm{Tr}(n n^t)^2 =\beta_1 [\mathrm{Tr}(T T^t) +\mathrm{Tr}(A A^t)+\mathrm{Tr}(S S^t)]^2$, hence it will not mix
the three sectors of $J=0,1$ and $2$.
Nevertheless, the non-linearity of the $\beta_2$-term will  mix them. 
Below we will study the ground state order parameter configuration in the cases of $\Delta \alpha_0>0$ and $\Delta \alpha_0<0$, respectively.

\subsubsection{Explicit SO symmetry breaking with 
$\Delta \alpha_0>0$ }

\begin{figure}[t]
\centering
\includegraphics[width=1.0\linewidth]{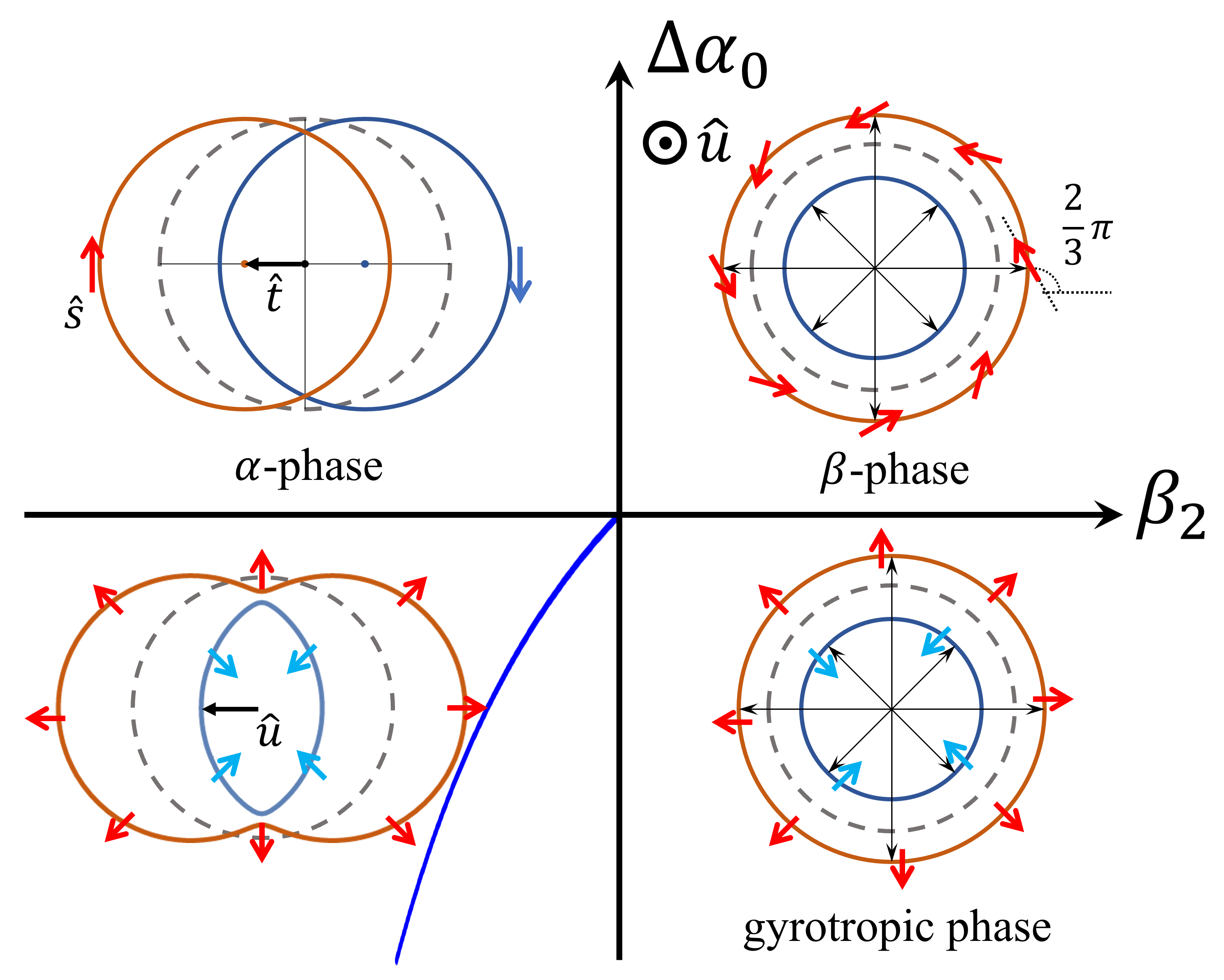}
\caption{Schematic phase diagram illustrating the  phases as a function of $\beta_2$ and $\Delta \alpha_0$.
Here, only the cross-section within the $k_x$-$k_y$ plane is depicted.
In the regime where $\Delta\alpha_0>0$, the system stabilizes in the $\alpha$-phase when $\beta_2<0$, characterized by a circular Fermi surface shifted along the $\hat{t}$-direction and spin polarization along $\hat{s}$-direction.
However, for positive $\beta_2$, the $\beta$-phase becomes stable, with its specific spin texture depending on $\Delta\alpha_0$.
If $\Delta\alpha_0>0$, the spin configuration results from a $2\pi/3$ rotation around the $\hat{u}$ axis to the reference configuration (e.g., the $J=0$ gyrotropic phase).
Alternatively, a gyrotropic phase is stabilized when $\Delta\alpha_0<0$, 
and this phase also persists for negative $\beta_2$ of small magnitude.
For $\beta_2<0$ and large enough in the magnitude, the multipolar component is induced and the Fermi surface of the opposite spin sector will distort in the opposite direction. 
}
\label{fig:3DpAlpha0Beta2}
\end{figure}

For $\Delta\alpha_0>0$, the minimization of the GL free energy favors the solution satisfying the traceless condition $\mathrm{Tr}n=0$.
In other words, the order parameter $n_{\mu b}$ prefers to project out the component of $J=0$ and keep the components of $J=1$ and $2$.
The phase diagram and the corresponding spin-momentum configurations around the  Fermi surface are shown in Fig.~(\ref{fig:3DpAlpha0Beta2}).

Let us consider the effect of a positive $\Delta \alpha_0$ in the $\beta$-phase, {\it i.e.}, in the case of $\beta_2>0$.
According to the result in Sect. \ref{sect:GL1}, the $\beta_2$-term reaches 
the minimum when $n n^t$ is a constant matrix
provided fixing $\mathrm{Tr}(n n^t)=\sum_i f_i^2=\lambda^2$.
This means that $n=\bar{n} R$ with $\bar{n}$ the magnitude of the order and $R$ is an $O(3)$ matrix.
Up to an inversion transformation, we assume that $R \in SO(3)$, which can be parameterized as $R(\hat n, \theta)$ with $\hat n\in S^2$ the rotation axis and $\theta$ the rotation angle. 
At $\theta=\frac{2}{3}\pi$, $\mathrm{Tr} (R)=1+2\cos\theta=0$.
As a result, a generic matrix form of 
$R(\hat u,\frac{2}{3}\pi)$ is defined as follows: For two 3-vectors $\mathbf{a}$
and $\mathbf{b}$, $R$ acts as
\begin{equation}
\mathbf{a} \cdot R  \cdot \mathbf{b} =
\frac{\sqrt{3}}{2} \mathbf{a} \cdot (\hat{u} \times \mathbf{b}) 
+\frac{3}{2} \left( (\mathbf{a} \cdot \hat{u})( 
\hat{u} \cdot \mathbf{b}) - \frac{1}{3} 
\mathbf{a} \cdot  \mathbf{b} \right),
\nonumber
\end{equation}
which is similar to the case of the Leggett angle found in the context of superfluid $^3$He-B \cite{leggett1975he3}.
The effective SOC becomes
\begin{equation}
\begin{aligned}
H_{\text{SOC}}= & \bar{n}\left[
\frac{\bm{\sigma}\cdot (\hat{u}\times \hat{\mathbf{k}})}{\sqrt{3}} 
+(\bm{\sigma}\cdot \hat{u}) ( \hat{u} \cdot \hat{\mathbf{k}})
-\frac{1}{3} \bm{\sigma}\cdot \hat{\mathbf{k}}\right].
\end{aligned}
\end{equation}

The Goldstone manifold of the order parameter can be parameterized as 
\begin{eqnarray}
T R(\hat u, \frac{2}{3}\pi) T^{-1} =
R(T\hat u, \frac{2}{3}\pi),
\end{eqnarray}
where $T$ is an $SO(3)$ matrix. 
When $T$ is the rotation around the axis of $\hat u$,
$R$ does not change, hence, the Goldstone manifold is
$SO\left(3\right)_{J}/SO\left(2\right)_{J}=S^{2}$ with two branches of Goldstone modes. 
For a particular example of $\hat{u}=\frac{1}{\sqrt{3}}(1,1,1)$, which results in the mean-field Hamiltonian,
\begin{equation}
H_{\text{SOC}}= \bar{n}\left(\sigma_{z}k_{x}+\sigma_{x}k_{y}+\sigma_{y}k_{z}\right).
\end{equation}
Hence, there is no non-trivial line defect since the
fundamental homotopy group $\pi_1(S^2)= 1$.
Nevertheless, there exist topologically non-trivial point defects as classified by the 2nd homotopy group
$\pi_2(S^2)=Z$.

Now consider the effect of a positive $\Delta \alpha_0$ to the $\alpha$-phase, i.e., the regime of $\beta_2<0$.
The ground state order parameter configuration is parameterized by $n=\bar{n} R$ with $R(\hat s, \hat t)$ defined as follows: For any two 3-vectors 
$\mathbf{a}$ and $\mathbf{b}$, $R$ acts as
\begin{equation}
\mathbf{a} \cdot R \cdot \mathbf{b} = 
(\mathbf{a} \cdot \hat s) (\hat t \cdot \mathbf{b}),
\end{equation}
where $\hat t$ and $\hat s$ are two unit vectors orthogonal to each other. 
It is easy to show that in this case 
$\mathrm{Tr}n=0$.
The effective SOC takes the following form,
\begin{equation}
\begin{aligned}
& H_{\text{SOC}} = -\bar{n}
(\hat{s}\cdot\bm{\sigma}) (\hat{t}\cdot\hat{\mathbf{k}}) .
\end{aligned}
\end{equation}
The two spherical Fermi surfaces shift in the direction of $\hat{t}$ with spin direction along $\hat{s}$, exhibiting a SOC that is a superposition of Rashba and Dresselhaus-like with equal weight. 

The residual symmetry of the $\alpha$-phase at $\Delta \alpha_0>0$ is the rotation around the axis of $\hat s \times \hat t$ at the angle of $\pi$.
Hence, the Goldstone manifold is:
\begin{eqnarray}
SO(3)_J/Z_2=SU(2)/Z_4=S^3/Z_4,
\end{eqnarray}
where $Z_4$ represents the $SU(2)$ rotation around $\hat s \times \hat t$ by the angle of $0,\pm \pi, 2\pi$.
Hence, the fundamental group of the Goldstone manifold is
$\pi_1 (SO(3)_J/Z_2)=Z_4$, which means that there exist 4 different types line defects characterized by the winding numbers of $0, \pm \frac{1}{2}, 1$.
There exist three branches of Goldstone modes, which correspond to rotations around three axes of $\hat t$, $\hat s$, and $\hat s \times \hat t$. 
The 2nd homotopy group  $\pi_2 (SO(3)_J/Z_2)=0$, 
which means that there is no non-trivial point defect.

\subsubsection{Explicit SO symmetry breaking with $\Delta \alpha_0<0$ }

The situation that $\beta_2>0$ favors the $\beta$-phase, and the gyrotropic state ($J=0$) favored by $\Delta \alpha_0<0$ 
also belongs to the $\beta$-phase. 
Hence, at $\beta_2>0$, the ground state takes the 
gyrotropic state, {\it i.e.}, $n_{\mu a}=\bar{n} \delta_{\mu a}$.
This is a $Z_2$ symmetry breaking phase, {\it i.e.}, parity breaking, but the $SO(3)_J$ symmetry is maintained. 

The situation of $\beta_2<0$ is more complicated since it favors the $\alpha$-phase which is incompatible with the gyrotropic state . 
Naturally, there exist a competition between them. 
This gyrotropic phase with effective SOC $H_{\text{SOC}}= \bar{n} \vec{\sigma} \cdot \hat{\mathbf{k}}$ is robust against a negative $\beta_2$ with a small magnitude.
In contrast, a negative $\beta_2$-term with a sufficiently large magnitude will mix the gyrotropic order.  
As far as the $\beta_2$-term is not so large, the ground state configuration will be given by $f_1\geq f_2=f_3$;
see Appendix \ref{appendixGL3d} for details based on the singular value decomposition (SVD).

The mixed phase on the left of the phase boundary exhibit an effective spin-orbit coupling taking for the form
\begin{align}
H_{\text{SOC}} =\bar{n} ( \bm{\sigma}\cdot \hat{\mathbf{k}}
+a (\bm{\sigma}\cdot \hat{u}) (\hat{\mathbf{k}}\cdot \hat{u})),
\label{eq:aniGyro}
\end{align}
where $a$ is the coefficient of the $\alpha$-phase component.
The two unit vectors representing the orbital and spin polarizations are parallel to each other in the $\alpha$-phase marked as $\hat u$ to maximize $\mathrm{Tr}(n n^t)$.
The Fermi surface configurations are marked in Fig.~(\ref{fig:3DpAlpha0Beta2}).
The residue symmetry of such a state is the
$SO(2)$ rotation around $\hat u$ and the $\pi$ angle rotation around any axis perpendicular to $\hat u$.
Hence the Goldstone manifold is
\begin{eqnarray}
SO(3)/D(\infty)=\mathrm{RP}^2, 
\end{eqnarray}
and its fundamental group is
$\pi_1(\mathrm{RP}^2)=Z_2$, which means that there exist two types of line defects characterized by
the winding numbers of $0$ and $\frac{1}{2}$.

\section{D-wave unconventional magnetism in 2D}
\label{sect:dwave}

In this section, we investigate unconventional magnetism within a 2D system, 
with a particular focus on the $l=2$ partial-wave ($d$-wave) channel.
Although the extension to higher partial-wave channels is straightforward, the $d$-wave case, characterized by its anisotropic nature and a winding number of $2$, provides a clear framework for elucidating the underlying physics.
The emergent phases in this channel can arise intrinsically from Pomeranchuk instabilities in Fermi liquids \cite{wu2007fermi,qian2025fragile} or be induced by lattice spin configuration \cite{smejkal2022beyond,smejkal2022emerging}.

\subsection{The d-wave order parameters}
Following the formalism set up in Sect. \ref{sect:GL1}, the order parameter for $d$-wave ($l=2$) unconventional magnetism is defined via the expectation value:
\begin{align}
n_{\mu b} \equiv \langle \hat{Q}_{\mu b}(\vec{r})\rangle 
= \langle \psi^{\dagger}(\vec{r}) \sigma_{\mu} g_b^{(2)}(-i\hat{\nabla}) \psi(\vec{r}) \rangle,
\label{eq:dwave_n}
\end{align}
where $g_{b}^{(2)}(\mathbf{k})$  are the basis functions for the $l=2$ channel in 2D $(b=1,2)$.
These basis functions possess a winding number of $2$ and are explicitly given by \cite{wu2007fermi}:
\begin{equation}
g_{1}^{(l=2)}(\mathbf{k})=\cos 2\theta_{\mathbf{k}},\quad
g_{2}^{(l=2)}(\mathbf{k})=\sin 2\theta_{\mathbf{k}} ,
\end{equation}
where $\theta_{\mathbf{k}}$ is the azimuthal angle of $\mathbf{k}$. 
The development of a non-zero order parameter $n_{\mu b}$ induces an effective momentum-dependent field, leading to the following term in the Hamiltonian:
\begin{equation}
\begin{aligned}
H_{d}=& \sigma_{x} \big( n_{x1} \cos2\theta_{\mathbf{k}} + n_{x2} \sin2\theta_{\mathbf{k}} \big)  \\
+& \sigma_{y} \big( n_{y1} \cos2\theta_{\mathbf{k}} + n_{y2} \sin2\theta_{\mathbf{k}} \big) \\
+& \sigma_{z} \big( n_{z1} \cos2\theta_{\mathbf{k}} + n_{z2} \sin2\theta_{\mathbf{k}} \big).
\end{aligned}
\end{equation}
For convenience, we use $\mu=x,y,z$ to represent the spin channel index.
The Hamiltonian explicitly breaks the TR symmetry, 
but preserves the inversion (parity) symmetry.


To elucidate the symmetry constraints on the $d$-wave unconventional magnetic order, we examine the transformation properties of $n_{\mu b}$ under simultaneous rotations in spin and orbital spaces. 
Consider a rotation by the angle $\theta$ about the $z$-axis. 
The $d$-wave basis functions transform as
\begin{equation}
(\cos2\theta_{\mathbf{k}},\sin2\theta_{\mathbf{k}})
\rightarrow
(\cos2(\theta_{\mathbf{k}}+\theta),\sin2(\theta_{\mathbf{k}}+\theta)).
\end{equation}
Under this operation, the $3\times 2$ order parameter matrix transforms as
\begin{equation}
n_{\mu b} \rightarrow  
R_{S,\mu\nu}(\theta) n_{\nu a}R_{L,ab}^{-1}(2\theta),
\end{equation}
where the rotation matrices in the spin and orbital sectors are given explicitly by,
in the spin space
\begin{equation*}
R_{S}(\theta) 
=\begin{pmatrix}
\cos\theta & \sin\theta & 0 \\
-\sin\theta & \cos\theta & 0    \\
0 & 0 & 1
\end{pmatrix},
\end{equation*}
and in the orbital space,
\begin{align*}
R_{L}(2\theta)
=\begin{pmatrix}
\cos 2\theta & \sin 2\theta\\
-\sin 2\theta & \cos 2\theta
\end{pmatrix}.
\end{align*}

\subsection{Spin-orbit coupling splitting}
In the presence of explicit SOC, the system symmetry is reduced to $O(2)_J$, which is the semi-direct product between the $SO(2)_J$ and mirror reflection with respect to a vertical plane. 
The order parameters may be decomposed into irreducible representations of the conserved $O(2)_J$ symmetry group.
We define the following combinations: 
\begin{eqnarray}
T_{1}&=&(n_{x1}+n_{y2},n_{y1}-n_{x2}) \nonumber \\
T_{2}&=&(n_{z1},n_{z2}),  \nonumber \\
T_{3}&=&(n_{y2}-n_{x1},n_{x2}+n_{y1}). 
\end{eqnarray}
$T_l$ at $l\neq 0$ consist of two bases, and their complex superposition carries the eigenvalue of $J_z=\pm l$.
Under the rotation angle $\theta$, they transform as
\begin{eqnarray}
T_{l} \rightarrow T_{l} R_{L}(l\theta).
\end{eqnarray}

The GL free energy at the quadratic order can be constructed as follows:
\begin{equation}
\begin{aligned}
& F_{2}[n] 
=\alpha_{1}T_{1}\cdot T_{1}+\alpha_{2}T_{2}\cdot T_{2}+\alpha_{3}T_{3}\cdot T_{3},\\
&=  \alpha_{1}[(n_{x1}-n_{y2})^{2}+(n_{x2}+n_{y1})^{2}]
\\
&+\alpha_{2}[n_{z1}^{2}+n_{z2}^{2}]  \\
& +\alpha_{3}[(n_{y1}-n_{x2})^{2}
+(n_{x1}+n_{y2})^{2}],
\end{aligned}
\label{eq:GL2Dquad}
\end{equation}
where the phenomenological parameters $\alpha_{J}$ ($J = 1,2,3$) depend on microscopic details such as temperature and band structure. 
The sign and magnitude of these parameters determine the relative stability of each phase. 
The minimization of this free energy selects the dominant $T_J$ component, thereby determining the realized $d$-wave phase.

\begin{figure}[t]
\centering
\includegraphics[width=1.0\linewidth]{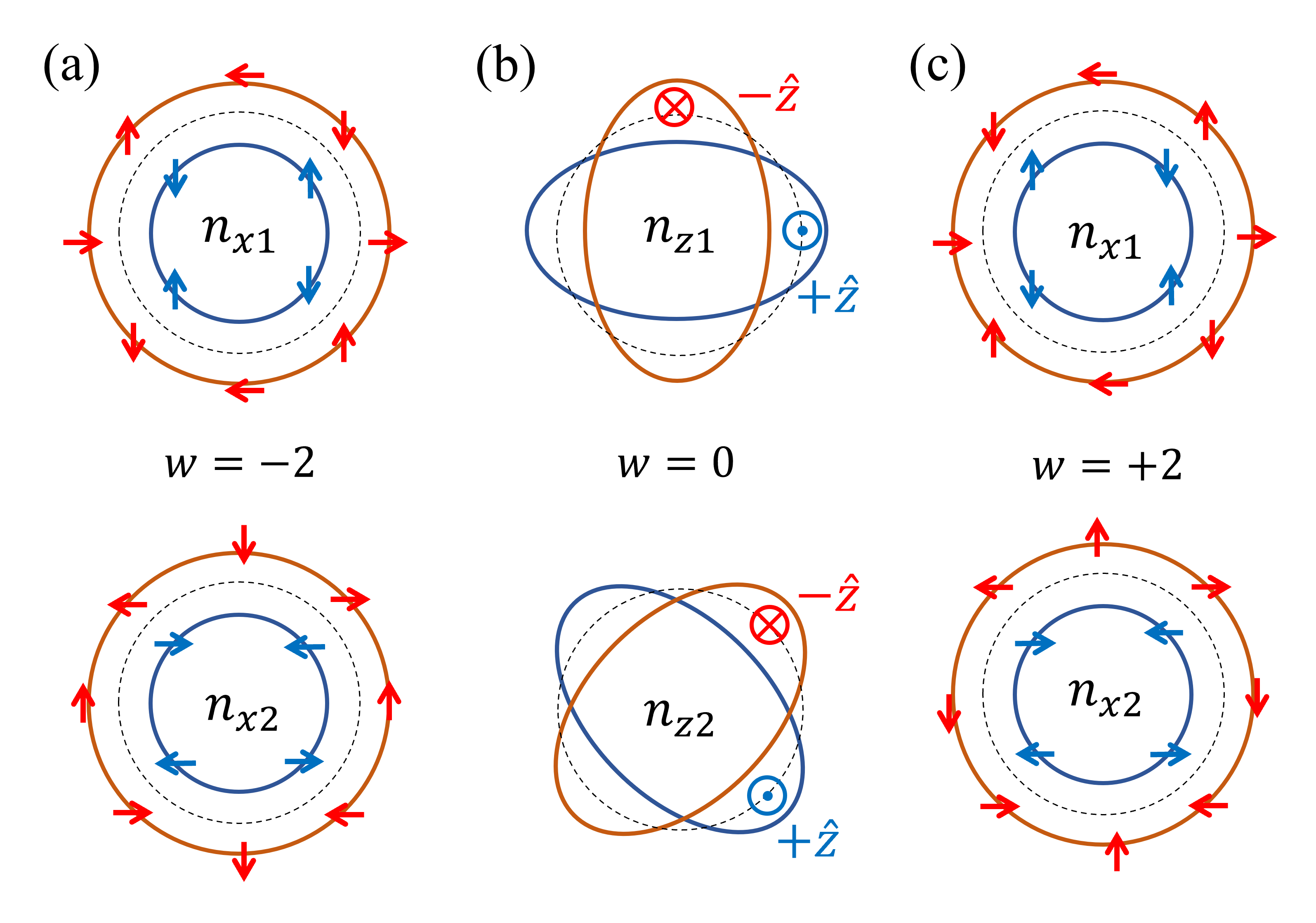}
\caption{Spin configurations in momentum space around the Fermi surface for the three channels: (a) $J_z =1$, (b) $J_z=2$, and (c) $J_z=3$.
In the presence of $d$-wave magnetism, the original Fermi surfaces (dotted circle) split into two Fermi surfaces (blue and red circles) with opposite spin polarizations.}
\label{fig:AlterTexture}
\end{figure}

Each of $T_{1,2,3}$ type of orders exhibit a distinct spin-orbit texture pattern over the Fermi surface. 
Despite the breaking of TR symmetry, the net magnetization vanishes. 
For the ordering in the $T_1$ channel, $n_{x1}=-n_{y2}$ and $n_{y1}=n_{x2}$. 
The corresponding effective Hamiltonian takes the form,
\begin{equation}
\begin{aligned}
H_{1} & =  n_{x1} \big(\sigma_{x} \cos2\theta_{\mathbf{k}}- \sigma_{y} \sin2\theta_{\mathbf{k}}\big)   \\
 &+ n_{x2} \big(\sigma_{x} \sin2\theta_{\mathbf{k}} +\sigma_{y} \cos2\theta_{\mathbf{k}} \big).
\end{aligned}
\end{equation}
Such an effective Hamiltonian results in a spin-splitting Fermi surface.
The spin orientation exhibits an anti-vortex structure with a winding number of $w=-2$ as shown in Fig.~\ref{fig:AlterTexture}($a$).
As for the $T_2$ channel, the effective Hamiltonian simplifies to
\begin{equation}
H_{2} = \sigma_{z} \big(n_{z1} \cos2\theta_{\mathbf{k}}+ n_{z2}\sin2\theta_{\mathbf{k}} \big).
\end{equation}
where $n_{z1}$ and $n_{z2}$ correspond to the $d_{x^2-y^2}$ and $d_{xy}$ configuration, respectively, as depicted in Fig.~\ref{fig:AlterTexture}($b$).
This phase, identified with the $\alpha$-phase \cite{wu2007fermi} or the ``altermagnetic" state \cite{smejkal2022beyond,smejkal2022emerging}, is characterized by a splitting of the originally spin-degenerate Fermi surface into two orthogonal ellipses with opposite spin polarization. 
The spin configuration in this case is topologically trivial.
Finally, for the $J=3$ channel,
$n_{x1}=n_{y2}$, $n_{x2}=-n_{y1}$, and $n_{z1}=n_{z2}=0$.
It leads to the effective Hamiltonian
\begin{equation}
\begin{aligned}
H_{3}= &n_{x1} \big(\sigma_{x} \cos2\theta_{\mathbf{k}} +\sigma_{y} \sin2\theta_{\mathbf{k}} \big)     \\
& +n_{x2}\big(\sigma_{x} \sin2\theta_{\mathbf{k}} -\sigma_{y} \cos2\theta_{\mathbf{k}} \big) .
\end{aligned}
\end{equation}
In this phase, the spin texture forms a vortex configuration in momentum space, yielding a winding number $w=+2$ as shown in 
Fig.~\ref{fig:AlterTexture}($c$).
The contrasting vortex and anti-vortex textures in the $J=3$ and $J=1$ channels, respectively, reflect the additive ($J=2+1$) versus subtractive ($J=2-1$) nature of the underlying angular momentum couplings.

The above orders exhibit the so-called ``spin-group" type symmetry. 
For the cases of that exhibits the winding number $w=\pm 2$,
there appears an effective 
$J^\prime_z=L_z+\frac{w}{2}\sigma_z$. 
In the case of $w=0$, {\it i.e.}, $J_z=2$, the residual symmetry is $SO(2)$ rotation around the $z$- axis in the spin channel, the $C_2$ rotation at the angle of $\pi$ in the orbital channel, and a $Z_2$ rotation with the angle of $\pi$ around an axis perpendicular to the spin axis.
Nevertheless, this kind of symmetries do not belong to the original $SO(2)_J$, hence, they can be viewed as emergent symmetries at low energies. 

All the above orders in the three channels of $T_{1,2,3}$ break the continuous rotational symmetry in the $J_z$ channel.
Since the total symmetry is $SO_J(2)$, the Goldstone manifold is also $SO_J(2)$. 
It gives rise to a single branch of gapless Goldstone modes, which are anticipated to dominate the low-energy dynamics of these $d$-wave magnetic states.

\subsection{The effect of the quartic $\beta_2$ terms}

We now incorporate quartic terms into the GL free energy and study their effect on the $d$-wave magnetism. 
Building upon the quadratic analysis above, we consider the full GL functional up to fourth order. 
For simplicity, we focus on the competition driven by $\alpha_2=\alpha+\Delta\alpha_2$ and $\beta_2$, assuming $\alpha_1=\alpha_3=\alpha$ remain negative. 
The free energy takes the form
\begin{equation}
\begin{aligned}
F[n]  =& \alpha \mathrm{Tr} (n n^t)
+\Delta\alpha_2 \big(n_{z1}^{2}+n_{z2}^{2}\big) \\
+& \beta_1 \big[\mathrm{Tr}(nn^t)\big]^2 
+ \beta_{2} \mathrm{Tr}\big[(n n^{t})^{2}\big].
\end{aligned}
\end{equation}
where $\mathrm{Tr}(n n^t) = |\vec{n}_1|^2 + |\vec{n}_2|^2$.
For convenience, we use the two-vector representation defined as
$\vec{n}_1 = (n_{x1}, n_{y1}, n_{z1})$ and $\vec{n}_2 = (n_{x2}, n_{y2}, n_{z2})$.
The quadratic $\beta_1(\mathrm{Tr}[(n^{t}n)])^{2}$ term primarily affects the magnitude of the order parameter, leading to the stability of the ordered state when $\alpha<0$ and $\beta_1>0$.
The $\beta_{2}$-term further influences the relative orientation of the spin vectors $\vec{n}_1$ and $\vec{n}_2$, which can be rewritten in the form \cite{wu2007fermi},
\begin{equation*}
\begin{aligned}
&\beta_{2}\mathrm{Tr}[(nn^{t})^{2}]
=\beta_{2}
\big(|\vec{n}_{1}|^{2} +|\vec{n}_{2}|^{2}\big)^{2} 
-2\beta_{2}
|\vec{n}_{1}\times\vec{n}_{2}|^{2}.
\end{aligned}
\end{equation*}
We analyze the ground state configuration based on the signs and magnitudes of $\Delta\alpha_2$ and $\beta_2$.

\subsubsection{$d$-wave unconventional magnetism with $\Delta \alpha_2>0$}

When $\Delta\alpha_2 > 0$, the $\Delta\alpha_2 \big(n_{z1}^{2}+n_{z2}^{2}\big)$-term disfavors the $J_z = 2$ component. 
The order parameters in terms of 
$\vec{n}_1$ and $\vec{n}_2$ 
tend to project out their $z$-component in spin orientation. 
The $\beta_2$ term further selects the specific ground state configuration.

\begin{figure}[t]
\centering
\includegraphics[width=1.0\linewidth]{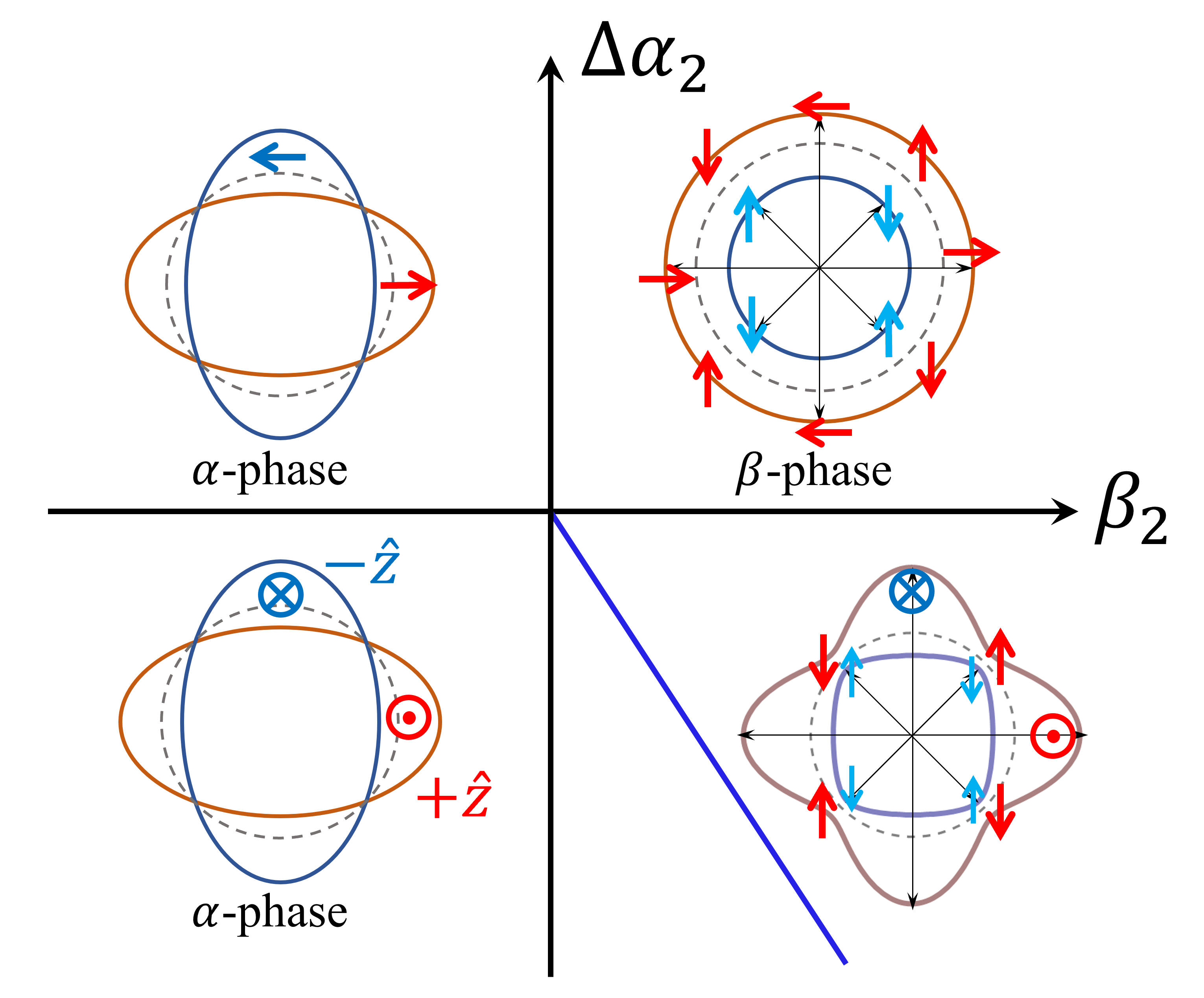}
\caption{Schematic phase diagram in the $\beta_2$-$\Delta\alpha_2$ parameter space, including representative configurations for the distinct phases.
The arrows indicate the in-plane spin texture.
For $\beta_2<0$ and $\Delta\alpha_2>0$,
the system favors an $\alpha$-phase characterized by parallel spin axes ($\vec{n}_1\parallel\vec{n}_2$) lying in the $xy$-plane;
in this phase, the circular Fermi surfaces split into two ellipsoids with associated in-plane spins pointing in opposite directions. 
When $\beta_2>0$ and $\Delta\alpha_2>0$, the favored state is the $\beta$-phase, featuring in-plane orthogonal spins of equal magnitude ($\vec{n}_1\perp\vec{n}_2$ and $|\vec{n}_1|=|\vec{n}_2|$) and a spin texture around the Fermi surface with winding number $w=\pm 2$.
In the region where $\beta_2<0$ and $\Delta\alpha_2<0$, a pure $J_z=2$ $\alpha$-phase state (also known as $d$-wave $\alpha$-phase) is stabilized, distinguished by spins aligned along the $z$-axis.
Finally, the quadrant with $\beta_2>0$ and $\Delta\alpha_2<0$ exhibits competition: 
for small positive $\beta_2$, the $J_z=2$ state may persist, 
whereas for large positive $\beta_2$, a mixed state emerges featuring distorted Fermi surfaces and complex spin textures.}
\label{fig:2DAlpha2Beta2}
\end{figure}

If $\beta_2 < 0$, the term $-2\beta_2 |n_1 \times n_2|^2$ is minimized when $\vec{n}_1 \parallel \vec{n}_2$.
This corresponds to the $\alpha$-phase configuration. 
Given the constraint from $\Delta\alpha_2 > 0$, we expect in-plane parallel spins. 
Since the cases of different winding numbers $w=\pm 2$ are energetically the same, their superposition yields the $\alpha$-phase with an in-plane spin configuration.
Without loss of generality, we can choose $\vec{n}_1=\bar{n}\hat{x}$ and $\vec{n}_2=0$, where $\bar{n}$ is the magnitude.
This leads to the effective $d$-wave spin-orbit Hamiltonian,
\begin{align}
H_d = \bar{n}\cos(2\theta_{\mathbf{k}}) \sigma_x.
\end{align}
This state represents the $\alpha$-phase with purely in-plane components of $\vec n_1$ and $\vec n_2$, as depicted in Fig.~(\ref{fig:2DAlpha2Beta2}).

If $\beta_2 > 0$, minimizing the $\beta_2$-term leads to $\vec{n}_1 \perp \vec{n}_2$ with equal amplitudes, $|\vec{n}_1| = |\vec{n}_2|$.
This favors the $\beta$-phase configuration. 
A possible realization compatible with $\Delta\alpha_2>0$ (in-plane spin) is $\vec{n}_1=\bar{n}\hat{x}$ and $\vec{n}_2=\bar{n}\hat{y}$. 
The corresponding Hamiltonians are
\begin{align}
H_d = \bar{n} \big(\cos(2\theta_{\mathbf{k}}) \sigma_x 
\pm \sin(2\theta_{\mathbf{k}}) \sigma_y\big).
\end{align}
These states exhibit a characteristic spin texture winding twice around the Fermi surface ($w=\pm2$), indicative of the $\beta$-phase \cite{wu2007fermi}, shown in Fig.~(\ref{fig:2DAlpha2Beta2}).

The residual symmetry in both phases is the rotation at the angle of $180^\circ$ 
combined with a reflection with time-reversal symmetry. 
Hence, the Goldstone manifold is $SO(2)_J/Z_2$.
The point defect is classified by the fundamental group $Z$.

\subsubsection{$d$-wave unconventional magnetism with $\Delta \alpha_2<0$}

We now turn to the parameter regime of $\Delta\alpha_{2}<0$.
According to the GL free energy functional, the $\Delta \alpha_2$-term
energetically favors configurations that maximize the components of the order parameter along the $z$-axis in spin space. 
The ground state structure is further determined by the sign of $\beta_2$.

When $\beta_2<0$, it favors the spin to be collinear,  $\vec{n}_1 \parallel \vec{n}_2$.
This condition is compatible with the tendency favored by $\alpha_2 < 0$ for spins to align along the $z$-axis.  
Minimizing the GL free-energy leads to 
$\vec{n}_1 = n_{z1} \hat{z}$ and $\vec{n}_2 = n_{z2}\hat{z}$ being parallel. 
Without loss of generality, this configuration can be represented by $\vec{n}_1=\bar{n}\hat{z}$ and $\vec{n}_2=0$. 
The effective $d$-wave spin-orbital term is given by
\begin{align}
H_d = \bar{n} \cos(2\theta_{\mathbf{k}}) \sigma_z,
\end{align}
corresponding to the $d$-wave $\alpha$-phase \cite{wu2007fermi,smejkal2022beyond,smejkal2022emerging}.
The Fermi surface exhibits characteristic splitting as depicted in Fig.~(\ref{fig:2DAlpha2Beta2}).

Conversely, when $\beta_{2}>0$, the quartic $\beta_{2}$-term favors orthogonal spin vectors, $\vec{n}_1 \perp \vec{n}_2$, while the condition $\Delta\alpha_2<0$ favoring $z$-alignment for both vectors, hence, this creates a competition. 
A detailed analysis using singular value decomposition (SVD) for the $3\times 2$ order parameter matrix $n_{\mu b}$ is presented in Appendix \ref{appendixGL2d}.
It indicates that the $\alpha$-phase remains stable to the regime of weak $\beta_2$.
When $\beta_2$ is large enough, minimizing the GL free energy yielding an order parameter structure characterized by two singular values, $f_1> f_2> 0$, where the dominant left singular vector (associated with $f_1$) aligns with the $\pm\hat{z}$ direction.
As shown in Appendix \ref{appendixGL2d}, this minimum energy configuration, under appropriate basis choices, corresponds to an effective Hamiltonian of the form, 
\begin{equation}
H_d =  f_1 \cos(2\theta_{\mathbf{k}}) \sigma_z+
f_2 \sin(2\theta_{\mathbf{k}}) \sigma_y,
\end{equation}
where $f_1\geq f_2\geq 0$ are determined by minimizing the free energy.
This Hamiltonian represents a mixed phase: the $f_1$ term describes the $J_z = 2$ component favored by $\Delta \alpha_2<0$, 
while the $f_2$ term introduces an in-plane spin component ($\sigma_y$) coupled to the $g_2^{(2)}(\hat{\mathbf{k}}) = \sin 2\theta_{\mathbf{k}}$ basis function, arising due to the influence of $\beta_2 > 0$. 
This phase exhibits distorted Fermi surfaces and more complex spin textures compared to the pure $J_z = 2$ state, as illustrated schematically in Fig.~(\ref{fig:2DAlpha2Beta2}). 
The precise determination of the magnitudes $f_1$, $f_2$ and the location of the phase boundary between the pure $J_z = 2$ state ($f_2 = 0$) and the mixed state ($f_2 > 0$) requires the detailed minimization of the free energy functional, as outlined in Appendix \ref{appendixGL2d}.

\section{Discussions}
\label{sect:dis}
Spin-orbit coupling always exists in real materials. 
SOC lowers the symmetry from two independent rotational symmetries in spin and orbital channels down to the overall rotation symmetry. 
Hence, in the study of unconventional magnetism, the role of SOC is similar to that of magnetic anisotropy in the study of magnetism.
As shown in Fig.~(\ref{fig:2DAlpha2Beta2}), the Fermi surface structures of the $\alpha$- and $\beta$-phase Fermi surface structures could still be maintained, but the relative configurations of momentum and spin are pinned by the external SOC.

The above analysis may bring caution when applying the concept of ``spin-group" in real materials, in particular, when the material contains heavy elements. 
Even though the Fermi surface splitting, or, the real space spin texture, exhibits spin-group type symmetry, it does not mean that SOC is weak. 
For example, the Fermi surface structures of Fig.~(\ref{fig:2DAlpha2Beta2}) exhibit the symmetry of an orbital rotation by the angle of $\frac{\pi}{2}$ followed by a spin rotation by the angle of $\pi$.
However, this symmetry applies only to the single-particle dispersion, and does not reflect the fundamental symmetry of the full Hamiltonian.
All properties beyond the single-particle level, such as collective excitations and topological structures, should not manifest spin-group properties.

The experimental exploration of unconventional magnetism in real materials has revealed a rich landscape of novel quantum phases, in which SOC plays an important role. 
Seminal examples like the transition-metal dichalcogenide semimetal $1$T-TiSe$_2$, the spin–orbit coupled metal Cd$_2$Re$_2$O$_7$ and  Sr$_2$(Ir,Rh)O$_4$ have served as crucial candidates for observing these phenomena. 
For instance, the detection of a gyrotropic phase in $1$T-TiSe$_2$ \cite{xu2020gyrotropic,ueda2021correlation} underscores the realization of chirality through spontaneous breaking of space-inversion, mirror-reflection, and rotoinversion symmetries in this material.  
In Cd$_2$Re$_2$O$_7$, investigations have uncovered a structural phase transition lacking conventional order \cite{harter2017parity,hiroi2018pyrochlore,hayami2019electric}, 
with evidence pointing towards a parity-breaking metallic phase driven by multipolar order parameters and enhanced spin-orbit coupling \cite{fu2015soc,harter2017parity}.
The intricate interplay of spin-orbit interactions and electron correlations further extends to unconventional Mott insulators like Sr$_2$(Ir,Rh)O$_4$, where hidden time-reversal symmetry breaking orders, potentially linked to novel topological effects, are being actively investigated \cite{jeong2017hidden}.

\section{Conclusions}
\label{sect:con}
In summary, we have systematically explored the possible symmetry-breaking patterns of unconventional magnetism
in the 3D $p$-wave and 2D $d$-wave channels under the explicit spin-orbit coupling. 
The GL free energies are constructed for exploring the phase diagram.
The spin-orbit symmetry splitting takes place at the quadratic order, which results in different channels classified based on the representation of the total angular momentum. 
In 3D, the $p$-wave unconventional magnetism exhibits channels with $J=0,1,2$, and in 2D the $d$-wave unconventional magnetism exhibits channels with $J_z=\pm 1,\pm 2, \pm 3$.
When the quartic terms were incorporated, we considered the simplified cases where two of the three channels are degenerate
at the quadratic level for both the 3D $p$-wave and 2D $d$-wave cases, respectively. 
Minimization of the GL free energies leads to the ground state configurations of the $\alpha$-phase, the $\beta$-phase, and mixtures between them. 
Furthermore, by tuning the parameters, we observe distinct distortions of the Fermi surfaces in both the $p$-wave and $d$-wave channels. 
These findings provide a deeper understanding of the interplay between unconventional magnetism and the explicit spin-orbit coupling which always exists in real materials.

\section*{Acknowledgments} 
We are grateful to the stimulating discussions with Xing Gu, Zhiming Ruan, Lunhui Hu and Zhuang Qian. 
C.W. is supported by the National Natural Science Foundation of China under the Grants No. 12234016 and No. 12174317.  
This work has been supported by the New Cornerstone Science Foundation.

\twocolumngrid


\appendix

\setcounter{equation}{0}
\setcounter{figure}{0}
\setcounter{table}{0}
\renewcommand{\thefigure}{A\arabic{figure}}

\renewcommand*{\theHfigure}{\thefigure}



\section{Details of the GL analysis in 3D}
\label{appendixGL3d}
In this appendix, we detail the minimization procedure for the Ginzburg-Landau (GL) free energy relevant to the $p$-wave channel ($l=1$) in the presence of spin-orbit coupling (SOC), 
focusing on the competition between terms favoring different types of order. 
As derived in the main text (cf. Eq.~\ref{eq:GLF2nSOC} and surrounding discussion) \cite{wu2007fermi}, the GL functional up to quartic order, including the SOC-induced splitting ($\Delta\alpha_0$), can be written by the $3\times 3$ order parameter matrix $n\equiv n_{\mu b}$ as:
\begin{equation}
\begin{aligned}
F[n] & =\alpha\, \mathrm{Tr}\big(n n^t\big) +\beta_1\, \big[\mathrm{Tr}(nn^t)\big]^2 \\
&+ \frac{\Delta\alpha_{0}}{3}\, \big(\mathrm{Tr}n\big)^{2}
+\beta_{2}\, \mathrm{Tr}\big[\big(n n^{t}\big)^{2}\big],
\end{aligned}
\label{eqApp:GL_Full}
\end{equation}
We are particularly interested in the parameter regime relevant to the competition discussed in Sect.~\ref{sect:pwave}, namely where $\alpha<0$, $\beta_1>0$, $\Delta\alpha_0<0$, and $\beta_2<0$. 
The term $\alpha<0$ drives the system into an ordered phase, while $\beta_1>0$ (and $\beta_1>|\beta_2|$) ensures stability by fixing the overall magnitude (norm) of the order parameter. 
The competition arises between the $\Delta\alpha_0<0$ term, which favors the isotropic gyrotropic ($J=0$) state, and the $\beta_2<0$ term, 
which favors the anisotropic $\alpha$-phase configuration.

To analyze the minimum of $F[n]$ for a general real $3\times 3$ matrix $n$, we employ the singular value decomposition (SVD):
\begin{align}
n= U\Sigma V^t,\quad
\Sigma=\begin{pmatrix}
f_1 &  & 	\\
& f_2 & \\
& & f_3
\end{pmatrix},
\label{eqApp:singularN}
\end{align}
where $U$ and $V$ are two $3\times 3$ orthogonal matrices ($U^tU=I$, $V^tV=I$),
and $f_1\geq f_2\geq f_3\geq 0$ are the singular values.  
Using the SVD, we can express the terms in the free energy:
\begin{align}
nn^t=U \begin{pmatrix}
f_1^2 &  & 	\\
& f_2^2 & \\
& & f_3^2
\end{pmatrix} U^t.
\end{align}
Notice that $f_i^2$ ($i=1,2,3$) are the eigenvalues of positive semi-definite symmetric real matrix $nn^t$, and $\mathrm{Tr}(nn^t)\equiv \lambda^2=\sum_if_i^2$.
The diagonalizability of $nn^t$ by an orthogonal matrix, ensured by the spectral theorem, forms the basis for the SVD of $n$.
In general, for any $3\times N$ matrix $n_{\mu b}$ with $N=2$ in 2D or $N=2l+1$ in 3D, $nn^t$ is always positive, semi-definite and symmetric.
The singular value decomposition takes similar form as Eq.~\ref{eqApp:singularN}, where the orthogonal matrices $V$ now becomes $N\times N$ component.

The free energy functional $F[n]$ can thus be written in terms of the singular values and the orthogonal matrices $U,V$:
\begin{equation*}
\begin{aligned}
F & =-|\alpha|\sum_i f_i^2
+|\beta_1| \big(\sum_i f_i^2\big)^2 \\
& -|\beta_2| \sum_i f_i^4 
+ \frac{\Delta\alpha_{0}}{3} \big(\mathrm{Tr}[W \Sigma]\big)^{2},
\end{aligned}
\end{equation*}
where $W=V^tU$ is also an orthogonal matrix.
We seek the configuration $(f_i, W)$ that minimizes $F$. 
Let us analyze the term involving $\Delta \alpha_0$. 
Since we consider $\Delta \alpha_0 < 0$, minimizing $F$ requires maximizing the term
\begin{align*}
\big( \mathrm{Tr}n \big)^2 
=\big( \mathrm{Tr}[W\Sigma] \big)^2 
= \Big( \sum_i W_{ii} f_i \Big)^2.
\end{align*}
The trace $|\mathrm{Tr}(W\Sigma)| = \left| \sum_i W_{ii} f_i \right| \leq \sum_i |W_{ii}| f_i$. 
Since $W$ is orthogonal, $|W_{ii}| \leq 1$. 
The maximum possible value of $|\mathrm{Tr}(W\Sigma)|$ for fixed $f_i$ occurs when $W_{ii} = 1$ for all $i$, which implies $W = I$. 
The condition $W = V^T U = I$ means $U = V$. 
In this case, the SVD becomes $n = V \Sigma V^T$, indicating that $n$ is a symmetric matrix, and the singular values $f_i$ are the absolute values of the eigenvalues of $n$. 
Therefore, the minimum of the free energy occurs within the manifold of symmetric matrices $n$.

\begin{figure}[t]
\centering
\includegraphics[width=0.65\linewidth]{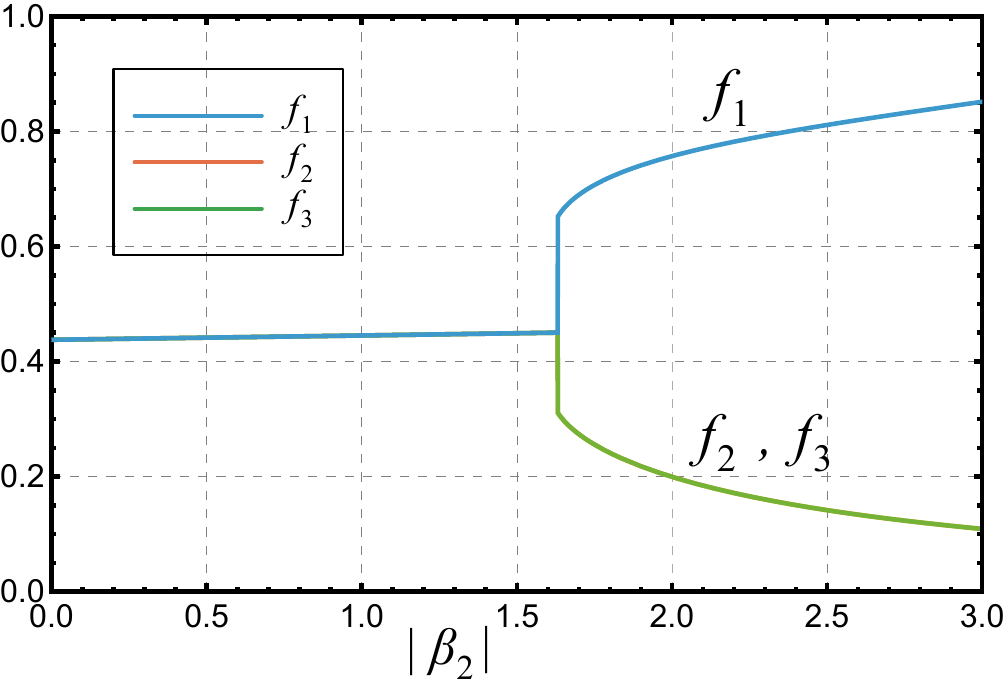}
\caption{Numerical minimization results for the singular values $(f_1, f_2, f_3)$ of the order parameter matrix $n$ (assuming $n$ is symmetric) as a function of $|\beta_2|$. Parameters used: $\alpha = -10$, $\beta_1 = 10$, $\Delta \alpha_0/ 3=-0.5$. 
For small $|\beta_2|$, the configuration is isotropic ($f_1 = f_2 = f_3$). 
As $|\beta_2|$ increases beyond a critical value, an anisotropic configuration with $f_1 > f_2 = f_3$ becomes energetically favorable.}
\label{fig:f1f2f3curve}
\end{figure}

Setting $U=V$, the free energy depends only on the singular values $f_i$ (which are non-negative here, representing the magnitude of eigenvalues):
\begin{equation}
\begin{aligned}
F\geq F_0[f_1, f_2, f_3] & \equiv \alpha \sum_i f_i^2 + \beta_1 \big(\sum_i f_i^2\big)^2 \\
& + \frac{\Delta\alpha_{0}}{3} \big(\sum_{i} f_i\big)^2 + \beta_{2} \sum_i f_i^4.
\end{aligned}
\label{eqApp:GL_Eigen}
\end{equation}
We now minimize $F_0$ in the regime $\alpha<0$, $\beta_1>0$, $\Delta\alpha_0<0$ and $\beta_2<0$.
The term $\beta_1(\lambda^2)^2+\alpha \lambda^2$ is minimized when $\lambda^2=|\alpha|/2\beta_1$, nearly fixing the overall scale.
The competition is between the $\Delta\alpha_0$ term and the $\beta_2$ term:
The term $\frac{\Delta \alpha_0}{3}(\sum f_i)^2$ (with $\Delta \alpha_0<0$) is minimized when $(\sum f_i)^2$ is maximized. 
For a fixed norm, $(\sum f_i)^2$ is maximized when $f_1=f_2=f_3$. 
This favors the isotropic gyrotropic state ($n \propto I$).
On the contrary, the term $\beta_2 \sum f_i^4$ (with $\beta_2 < 0$) is minimized when $\sum f_i^4$ is maximized. 
For a fixed norm, $\sum f_i^4$ is maximized when one eigenvalue dominates, i.e., $f_1 = f, f_2 = f_3 = 0$. 
This favors the anisotropic $\alpha$-phase ($n \propto \hat{u} \hat{u}^t$).

Consider the structure of the solution. 
The $\Delta \alpha_0$ term favors making the eigenvalues $f_i$ as close as possible. 
Let's compare a general state $(f_1, f_2, f_3)$ with one where the two smaller eigenvalues are averaged, $(f_1, \bar{f}, \bar{f})$ where $\bar{f}^2 = \frac{f_2^2 + f_3^2}{2}$, 
\begin{align*}
&F_0[f_1,f_2,f_3] - F_0[f_1,\bar{f},\bar{f}] \\
=& \frac{|\Delta\alpha_{0}|}{3} \Big(2 \sqrt{2} f_1 \sqrt{f_2^2+f_3^2}-2 f_1 (f_2+f_3) \Big) \\
+&\frac{|\Delta\alpha_{0}|}{3} (f_2-f_3)^2-\frac{|\beta_2|}{2} (f_2^2-f_3^2)^2
\end{align*}
Using $2(f_2^2 + f_3^2) \geq (f_2 + f_3)^2$, the term in the first line is always positive.
For the contribution in the second line, $\Delta\alpha_0$ would dominate if $|\beta_2|$ is smaller.
This suggests that the minimum energy configuration with smaller $|\beta_2|$ satisfies $f_2 = f_3 = \bar{f}$ (assuming $f_1 \geq f_2, f_3$).
In the following, we restrict to this regime.

Therefore, the ground state configuration corresponds to a symmetric matrix $n = V \Sigma V^T$ with $\Sigma=\mathrm{diag}(f_1, f_2, f_2)$, where the eigenvalues $f_1 \geq f_2 \geq 0$. 
The free energy becomes:
\begin{equation*}
\begin{aligned}
F_0[f_1,f_2,f_2]=& -|\alpha|(f_1^2+2f_2^2)
+|\beta_1| (f_1^2+2f_2^2)^2 \\
&-|\beta_2| (f_1^4+2f_2^4)
-\frac{|\Delta\alpha_{0}|}{3} (f_1+2f_2)^2
\end{aligned}
\end{equation*}
Minimizing this function with respect to $f_1$ and $f_2$ determines the ground state.
If $|\beta_2|$ is sufficiently small compared to $|\Delta\alpha_0|$, the $\Delta\alpha_0$-term dominates, favoring the isotropic solution $f_1=f_2$.
If $|\beta_2|$ is larger and involved, anisotropy will emerge with $f_1>f_2$.

\begin{figure}[t]
\centering
\includegraphics[width=0.8\linewidth]{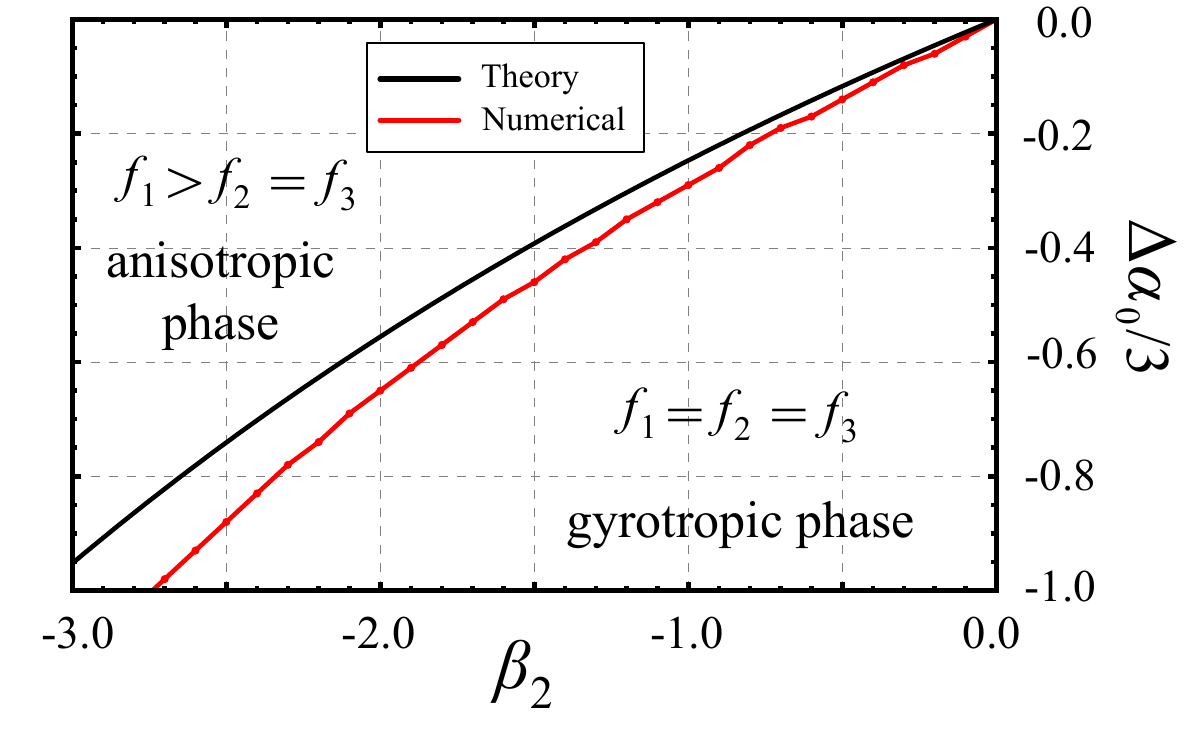}
\caption{Comparison of the theoretically derived critical boundary (black solid line) and the numerically determined phase boundary (red dashed line) in the ($\Delta \alpha_0 / 3$, $\beta_2$) plane between the isotropic configuration ($f_1 = f_2 = f_3$) favored for small $|\beta_2|$ and the anisotropic configuration ($f_1 > f_2 = f_3$) due to larger $|\beta_2|$. 
Parameters used: $\alpha = -10$, $\beta_1 = 10$. }
\label{fig:numcurve}
\end{figure}

The saddle point solution is given by $dF_0/df_1=dF_0/df_2=0$.
To determined the critical values of $\beta_2$ that marks the boundary between the isotropic and anisotropic phases for a given $\Delta\alpha_0$, we perform a stability analysis of the isotropic solution $f_i=f$ ($i=1,2,3$).
The value of $f$ in the isotropic state is found by minimizing $F_0[f,f,f]$, yielding
\begin{align*}
\frac{d }{d f} F_0[f,f,f]=0,\quad
f=\sqrt{\frac{|\alpha|+|\Delta\alpha_0|}{6 |\beta_1| -2 |\beta_2|}}
\end{align*}
We then investigate the stability of this isotropic solution against perturbations that break the symmetry towards the $f_1>f_2=f_3$ configuration.
A convenient perturbation representing this tendency is $f_1=f+2\delta$, $f_2=f_3=f-\delta$, which keeps the norm fixed to the linear order in $\delta$.
We expand the free energy difference for small $\delta$,
\begin{align*}
&F_0[f+2\delta,f-\delta,f-\delta]-F_0[f,f,f]  \\
=&-\frac{6 ( -3 |\Delta\alpha_0| |\beta_1| +2 |\alpha| |\beta_2| +3 |\Delta\alpha_0| |\beta_2|)}{3 |\beta_1|-|\beta_2|} \delta^2
+\mathcal{O}(\delta^3)
\end{align*}
The isotropic state (corresponding to $\delta=0$) is stable if the coefficient of the $\delta^2$ term above is positive (i.e. energy increases for $\delta\neq 0$).
The onset of instability, marking the emergence of the anisotropic phase ($\delta=0$), occurs when this coefficient becomes negative. 
The critical condition is found by setting the coefficient to zero, which gives
\begin{align}
|\beta_{2c}|= \frac{3 |\Delta\alpha_0| |\beta_1|}{2 |\alpha|+3 |\Delta\alpha_0|}.
\label{eq:beta2_critical}
\end{align}
This critical boundary is explicitly shown in Fig.~\ref{fig:f1f2f3curve}.
Exact analytic solution for the critical boundary can be obtained from considering the anisotropic saddle point.
There is no straightforward functional form for this critical boundary and we omit this for simplicity.

This stability analysis is corroborated by direct numerical minimization of the free energy functional $F_0[f_1, f_2, f_3]$, as shown in Fig.~\ref{fig:f1f2f3curve}. 
The results verify that the ground state is either isotropic ($f_1 = f_2 = f_3$) or anisotropic with two degenerate eigenvalues ($f_1 > f_2 = f_3$). 
Figure~\ref{fig:numcurve} explicitly compares the analytical phase boundary predicted by Eq.~\ref{eq:beta2_critical} with the boundary determined from the numerical minimization. 
The perturbative analysis gives rise the roughly order and tendency of the critical boundary compared to the numerical simulation. 
The plot illustrates that a larger anisotropy-driving term ($|\beta_{2}|$) is required to overcome the isotropy-favoring term ($|\Delta\alpha_0|$) when the latter is stronger.

In the anisotropic phase ($f_1 > f_2 = f_3$), the order parameter matrix $n = V \, \mathrm{diag}(f_1, f_2, f_2) V^T$ can be decomposed. 
Let $\hat{u}$ be the eigenvector corresponding to $f_1$ (the first column of $V$). 
Then
\begin{align*}
n &= V \begin{pmatrix} 
f_2 & & \\ & f_2 & \\ & & f_2 
\end{pmatrix} V^t 
+ V \begin{pmatrix} 
f_1-f_2 & & \\ & 0 & \\ & & 0 
\end{pmatrix} V^t \\
&= f_2 I + (f_1-f_2) \hat{u} \hat{u}^t.
\end{align*}
Substituting this into the effective SOC term $H_{\text{SOC}}=\sum_{\mu b} n_{\mu b} \sigma_\mu \hat{\mathbf{k}}^b$, and using the relation $ \sigma_\mu (\hat{u} \hat{u}^t)_{\mu b} \hat{\mathbf{k}}^b = (\sigma \cdot \hat{u})(\hat{\mathbf{k}} \cdot \hat{u})$, we obtain the effective Hamiltonian contribution:
\begin{align*}
H_{\text{SOC}}
=f_2 \bm{\sigma}\cdot\hat{\mathbf{k}}
+(f_1-f_2) (\bm{\sigma}\cdot\hat{u})
(\hat{\mathbf{k}}\cdot\hat{u})
\end{align*}
This Hamiltonian represents a superposition of the isotropic gyrotropic SOC ($J = 0$ type, favored by $\Delta \alpha_0 < 0$) and an anisotropic term characteristic of the $\alpha$-phase (favored by $\beta_2 < 0$), giving rise to the Eq.~\ref{eq:aniGyro} in the maintext.

\section{Details of the GL analysis in 2D}
\label{appendixGL2d}
The Landau functional up to the quartic order in the presence of $d$-wave unconventional magnetism is given by,
\begin{equation}
\begin{aligned}
F[n] & =\alpha\, \mathrm{Tr}\big(n n^t\big) +\Delta\alpha_2 \big(n_{z1}^{2}+n_{z2}^{2}\big) \\
&+\beta_1\, \big[\mathrm{Tr}(nn^t)\big]^2
+\beta_{2}\, \mathrm{Tr}\big[\big(n n^{t}\big)^{2}\big],
\end{aligned}
\end{equation}
or in the form
\begin{equation*}
\begin{aligned}
&F[\vec{n}_1,\vec{n}_2]
=\alpha(|\vec{n}_1|^2 +|\vec{n}_2|^2)
+\Delta\alpha_2 \big(n_{z1}^{2}+n_{z2}^{2}\big) \\
&+(\beta_{1}+\beta_2)
\big(|\vec{n}_{1}|^{2} +|\vec{n}_{2}|^{2}\big)^{2} 
-2\beta_{2}
|\vec{n}_{1}\times\vec{n}_{2}|^{2}.
\end{aligned}
\end{equation*}
For positive $\beta_1+\beta_2>0$ and negative $\alpha<0$, finite order parameter $n_{\mu b}$ is favored.
The structure of $n_{\mu b}$ or $(\vec{n}_1,\vec{n}_2)$ will be determined by $\Delta\alpha_2$ and $\beta_2$.

\begin{figure}[t]
\centering
\includegraphics[width=0.65\linewidth]{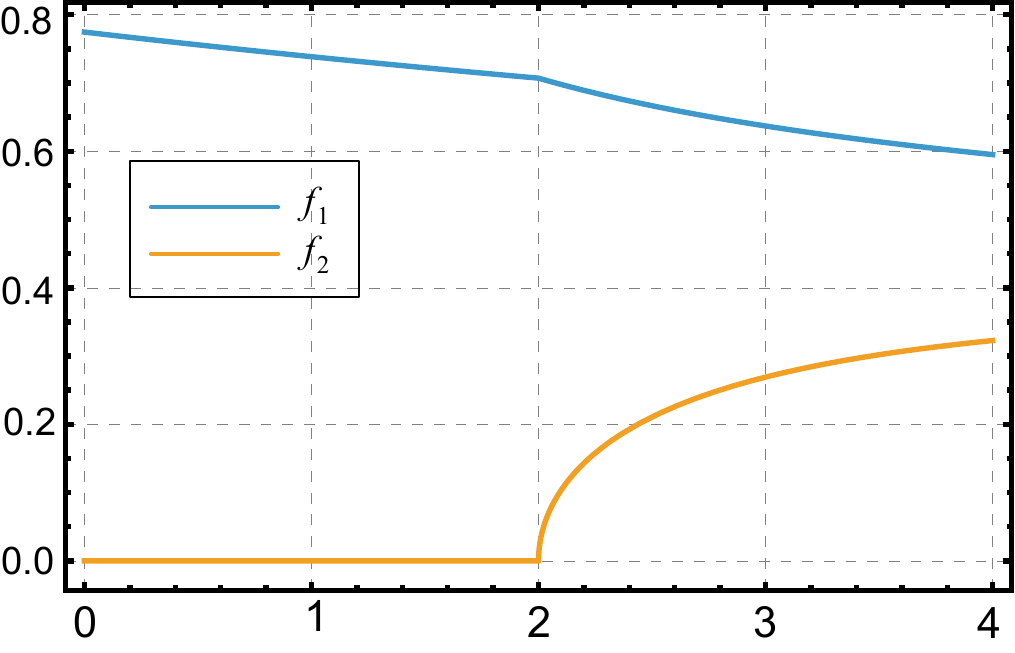}
\caption{Numerical minimization results for the singular values $(f_1, f_2)$ of the order parameter matrix $n$ (assuming $n$ is symmetric) as a function of $|\beta_2|$. Parameters used: $\alpha = -10$, $\beta_1 = 10$, $\Delta \alpha_2=-2$. 
For small $|\beta_2|$, the configuration is $f_2=0$. 
As $|\beta_2|$ increases beyond a critical value, $f_2$ becomes finite.}
\label{fig:f1f2curve2D}
\end{figure}

For $\Delta\alpha_2>0$, the spin vectors, $\vec{n}_1$ and $\vec{n}_2$, tend to locate within the plane.
Then, positive $\beta_2>0$ favors $\vec{n}_1\perp\vec{n}_2$ and $|\vec{n}_1|=|\vec{n}_2|$, e.g., we can choose $\vec{n}_1=\bar{n}\hat{x}$ and $\vec{n}_2=\bar{n}\hat{y}$.
For negative $\beta_2<0$, it will favor $\vec{n}_1\parallel\vec{n}_2$, and we can choose $\vec{n}_1=\bar{n}\hat{x}$ with $\vec{n}_2=0$.

For $\Delta\alpha_2<0$, the spin vectors, $\vec{n}_1$ and $\vec{n}_2$, tend to align within $z$-direction in the spin space.
For negative $\beta_2<0$, it will favor $\vec{n}_1\parallel\vec{n}_2$, and we can choose $\vec{n}_1=\bar{n}\hat{z}$ with $\vec{n}_2=0$.
Positive $\beta_2>0$ will induce frustration within $\Delta\alpha_2<0$.
This case will becomes more complicated.
When $\Delta\alpha_2<0$ dominate, small $\beta_2$-term is higher order in $n$, and the $\alpha$-phase is stable against weak $\beta_2$ perturbation.
To further clarify this picture, we consider the SVD as before.

For the real $3\times 2$ matrix $n_{\mu b}$ ($\mu=x,y,z=1,2,3$ and $b=1,2$), 
its singular value decomposition is given by,
\begin{align}
n= U\Sigma V^t,\quad
\Sigma=\begin{pmatrix}
f_1 & 0 	\\
0 & f_2 \\
0 & 0
\end{pmatrix},
\label{eqApp:singularN2D}
\end{align}
with $3\times 3$ orthogonal matrix $U$ and $2\times 2$ orthogonal matrix $V$ ($U^tU=I$, $V^tV=I$),
combined with real values $f_1\geq f_2\geq 0$.
Then, we have
\begin{align}
nn^t=U \begin{pmatrix}
f_1^2 &  & 	\\
& f_2^2 & \\
& & 0
\end{pmatrix} U^t,\quad
\mathrm{Tr}(nn^t)
=f_1^2+f_2^2,
\end{align}
and $f_i^2$ ($i=1,2$) are eigenvalues of the symmetric matrix $nn^t$.

\begin{figure}[t]
\centering
\includegraphics[width=0.7\linewidth]{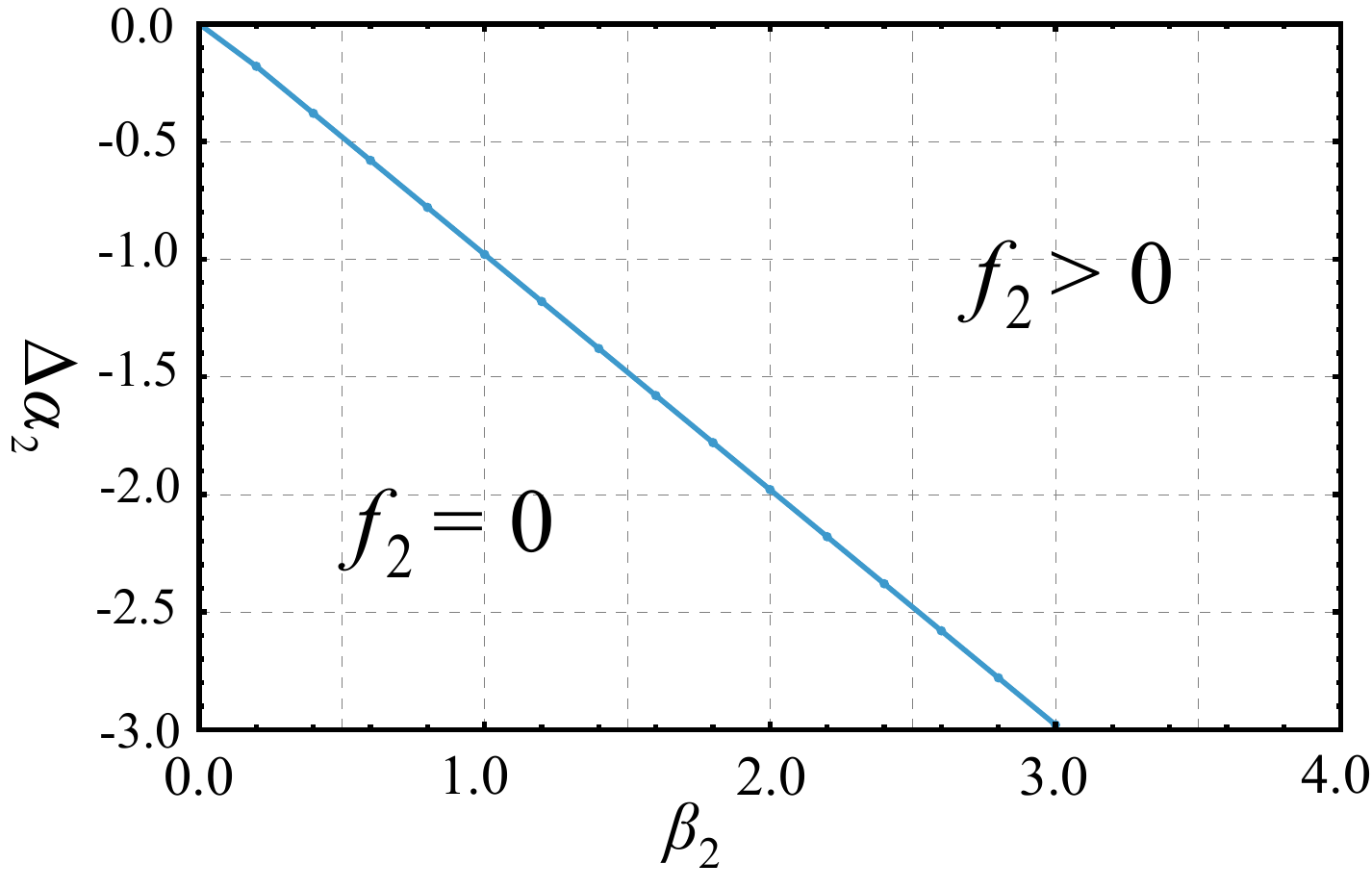}
\caption{Numerically determined phase boundary (orange dashed line) in the $\Delta \alpha_2-\beta_2$ plane between the $f_2=0$ configuration and $f_2>0$ configuration. 
Parameters used: $\alpha = -10$, $\beta_1 = 10$. }
\label{fig:numcurve2D}
\end{figure}

From the SVD, we can also have
\begin{align*}
n_{z1}^2 +n_{z2}^2
=U_{31}^2 f_1^2 +U_{32}^2 f_2^2
\leq f_1^2,
\end{align*}
where under the constraint $f_1\geq f_2$ and $U_{31}^2+U_{32}^2+U_{33}^2=1$, the above quantity is maximized when $U_{31}=\pm 1$.
Thus, we can have
\begin{align*}
F \geq F_0=&  -|\alpha|(f_1^2+f_2^2) +|\beta_1| (f_1^2+f_2^2)^2 \\
&-|\Delta\alpha_2| f_1^2 +|\beta_{2}| (f_1^4 +f_2^4),
\end{align*}
When $|\beta_2|=0$, $F_0$ favors $f_2=0$.
On the contrary, when $|\Delta\alpha_2|=0$, $F_0$ favors $f_1=f_2$.
The numerical simulation result is depicted in Fig.~\ref{fig:f1f2curve2D}.
The phase boundary between the $f_2=0$ state and $f_2>0$ state is numerically simulated, as shown in Fig.~\ref{fig:numcurve2D}.

Analytically, we can introduce the parameterization, $f_2/f_1=t$ and $f^2=f_1^2+f_2^2$, or inversely, $f_1=f/\sqrt{1+t^2}$ and $f_2=ft/\sqrt{1+t^2}$.
The saddle point is given by the equation $\partial F_0[f,t]/\partial f=0$, leading to
\begin{align*}
f^2 = \frac{|\alpha| (t^2+1) +|\Delta\alpha_2|}{(\beta_1+\beta_2)(t^4+1) +2\beta_1 t^2 } \frac{t^2+1}{2}\geq 0
\end{align*}
which is always possible.
For the stabilization of this solution, we consider $g(t)=\partial F_0[f,t]/\partial t=0$, namely,
\begin{align*}
t=0,\quad t^2=\frac{|\alpha|\beta_2 -|\Delta\alpha_2|\beta_1}{|\Delta\alpha_2|\beta_1 +(|\alpha|+|\Delta\alpha_2|)\beta_2}.
\end{align*}
The critical boundary is given by 
\begin{align*}
\beta_2= -\frac{\beta_1}{|\alpha|} \Delta\alpha_2,
\end{align*}
which matchess the numerical results in Fig.~\ref{fig:numcurve2D} exactly.

The generic ground state configuration in this case is given by,
\begin{align*}
n=& \begin{pmatrix}
\pm \hat{z} & \hat{u}_2 & \hat{u}_3
\end{pmatrix}\begin{pmatrix}
f_1 & 0 	\\
0 & f_2 \\
0 & 0
\end{pmatrix}
\begin{pmatrix}
\hat{v}_1^t \\ \hat{v}_2^t
\end{pmatrix} 
=\begin{pmatrix}
\pm f_1 \hat{z} & f_2 \hat{u}_2
\end{pmatrix}
\begin{pmatrix}
\hat{v}_1^t \\ \hat{v}_2^t
\end{pmatrix}.
\end{align*}
For precise, we can choose $\hat{u}_2=\hat{y}$ and $V=I$, 
which corresponds to the effective $d$-wave spin-orbital coupled term,
\begin{align*}
H_d= \pm f_1 \sigma_z \cos(2\theta_{\mathbf{k}})
+f_2 \sigma_y \sin(2\theta_{\mathbf{k}}).
\end{align*}
leading to the term considered in the maintext.

\twocolumngrid
\bibliography{SOCRef}

\end{document}